\documentclass{jfm}
\usepackage{graphicx}
\usepackage{epstopdf,epsfig}
\usepackage[normalem]{ulem}
\usepackage{amsmath,amssymb,mathtools,amsfonts,todonotes}
\usepackage{overpic,macros,natbib}
\usepackage{bm}
\usepackage{times}
\usepackage{gensymb}
\usepackage{textcomp}

\newcommand{\defeq}{\vcentcolon=}
\def \bcdot {\boldsymbol{\cdot}}

\shorttitle{A homogenised model for flow, transport and sorption in a heterogeneous porous medium}
\shortauthor{L.~C.~Auton, S.~Pramanik, M.~P.~Dalwadi, C.~W.~MacMinn, and I.~M.~Griffiths}

\title{A homogenised model for flow, transport and sorption in a heterogeneous porous medium}

\author{L.~C.~Auton\aff{1,2},
        S.~Pramanik\aff{2,3}\footnote{This author contributed significantly to this work.},
        M.~P.~Dalwadi\aff{1},
        C.~W.~MacMinn\aff{2},
   \and I.~M.~Griffiths\aff{1}\corresp{\email{Ian.Griffiths@maths.ox.ac.uk}}}

\affiliation{\aff{1}Mathematical Institute, University of Oxford, Oxford, OX2 6GG, UK
             \aff{2}Department of Engineering Science, University of Oxford, Oxford, OX1 3PJ, UK
             \aff{3}Department of Mathematics, Indian Institute of Technology Gandhinagar, Palaj, Gandhinagar -- 382355, Gujarat, India}

\begin{document}

\maketitle

\begin{abstract}
A major challenge in flow through porous media is to better understand the link between microstructure and macroscale flow and transport. For idealised microstructures, the mathematical framework of homogenisation theory can be used for this purpose. Here, we consider a two-dimensional microstructure comprising an array of obstacles of smooth but arbitrary shape, the size and spacing of which can vary along the length of the porous medium. We use homogenisation via the method of multiple scales to systematically upscale a novel problem involving cells of varying area to obtain effective continuum equations for macroscale flow and transport. The equations are characterised by the local porosity, a local anisotropic flow permeability, an effective local anisotropic solute diffusivity, and an effective local adsorption rate. These macroscale properties depend nontrivially on the two degrees of microstructural geometric freedom in our problem: obstacle size, and obstacle spacing. We exploit this dependence to construct and compare scenarios where the same porosity profile results from different combinations of obstacle size and spacing. We focus on a simple example geometry comprising circular obstacles on a rectangular lattice, for which we numerically determine the macroscale permeability and effective diffusivity. We investigate scenarios where the porosity is spatially uniform but the permeability and diffusivity are not. Our results may be useful in the design of filters, or for studying the impact of deformation on transport in soft porous media.
\end{abstract}

\begin{keywords} 
Homogenisation theory, porous media, solute transport, method of multiple scales, fluid dynamics, filtration. 
\end{keywords}

%%%%%%%%%%%%%%%%%%%%%%%%%%%%%%%%%%%%%%%%%%%%%%%%%%%%%%
\section{Introduction}
\label{s:intro}
%%%%%%%%%%%%%%%%%%%%%%%%%%%%%%%%%%%%%%%%%%%%%%%%%%%%%%

Fluid flow and solute transport in porous media occur in a wide variety of situations, including contaminant transport~\citep{quintard1994convection, brusseau1994transport}, lithium-ion batteries~\citep{li2018suppressing}, hydrogeological systems~\citep{domenico1998physical}, biofilms \citep{davit2013hydrodynamic}, bones \citep{fritton2009fluid}, and soils~\citep{daly2015homogenization}. Many of these porous media, including soils, rocks and biological tissues, are intrinsically heterogeneous and/or anisotropic at the pore scale, and macroscopic flow and transport in these systems are known to depend critically on pore structure and pore-scale fluid--solid interactions. For example, complex flow patterns and the resulting solute transport are believed to be crucial to the ecohydrology of peatlands, and have been attributed to the pore-scale heterogeneity and anisotropy of peat soil~\citep{beckwith2003anisotropy, wang2020effect}. \citet{clavaud2008permeability} used imaging to study the relationship between pore geometry and permeability anisotropy in sandstone, limestone, and volcanic rocks, finding that macroscopic flow properties depend on the details of the pore structure across these different rock types. \citet{o2015multiscale} use modelling in the context of tissue engineering to show that the microstructure induces anisotropy in flow properties, highlighting the role of microstructure in determining flow patterns and nutrient delivery. Changes in pore structure can also lead to large deviations from macroscopic models derived for homogeneous microstructures; for example, \citet{rosti2020breakdown} find that microstructural changes due to deformation of the solid skeleton can lead to a breakdown of Darcy's law. Ultimately, many aspects of the impacts of pore structure on macroscale flow and transport behaviour remain poorly understood. We focus here on the specific roles of pore-scale heterogeneity and anisotropy in the context of a simple, two-dimensional model problem.
 
Porous media are characterised by at least two distinct length scales: the characteristic length of each pore/solid grain (pore-scale) and the characteristic length of the porous medium itself (macroscale) \citep{tomin2016investigating}. Studying the impact of the pore structure on flow, transport and sorption via direct numerical simulation (DNS) in a complex geometry is computationally expensive, and can be prohibitively so when the pore-scale and macroscale lengths differ by orders of magnitude. For example, \citet{olivieri2020turbulence} used DNS to study turbulent flow through a cube containing randomly distributed solid fibres, considering up to 1000 fibers of length $1/2$ in a cube of side length $2\pi$. Similarly, \citet{kuwata2017direct} used DNS to study turbulent flow through a channel with a porous bed; the bed was four pores thick, with a square-frame structure. When there is a large number of obstacles or pores as would be relevant to practical applications, one way to deal with these disparate length scales is to systematically derive an upscaled macroscale model that is uniformly valid on the entire porous medium, and contains pertinent pore-scale information via the permeability, effective diffusivity, and an effective source/sink term.

There are many common methods for upscaling equations, including the method of moments, renormalisation group theory, and homogenisation via volume averaging or the method of multiple scales (MMS)~\citep{bensoussan2011asymptotic, mei2010homogenization, hornung1996homogenization,wood2003volume,salles1993taylor}. These different methods have been compared both with each other and with other DNS~\citep[\textit{e.g.},][]{salles1993taylor, davit2013homogenization, kuwata2017direct}. The two homogenisation methods yield the same macroscale equations, but via different routes. In essence, both methods identify the governing equations on the pore-scale, which are subject to closure conditions, and use this pore-scale problem to derive a system of equations over the macroscale. The formal nature of the MMS enables higher-order corrections to the leading-order macroscale equations to be determined.  Conversely, the volume-averaging method can be more physically intuitive \citep[see, for example,][]{wood2003volume, whitaker2013method,whitaker1986flow, davit2013homogenization} but it is more difficult to determine higher-order corrections and precise quantification of errors.

Classic homogenisation requires the microstructure to be strictly periodic at some scale. This requires a `periodic cell' for the MMS \citep{mauri1991dispersion, salles1993taylor, chapman2008multiscale, shipley2010multiscale} and a `representative elementary volume' for the volume-averaging method \citep{auriault1991heterogeneous,davit2013homogenization}. However, recent work has extended this technique to allow for slowly varying microstructure (\textit{i.e.}, microstructure that is only locally periodic) \citep[\textit{e.g.}][]{van2009crystal,van2011homogenisation,valdes2011volume,ray2012multiscale,muntean2020colloidal,richardson2011derivation,bruna2015diffusion,dalwadi2015understanding, dalwadi2016multiscale}. \citet{dalwadi2015understanding}, in particular, considered diffusive and advective transport through an array of impermeable obstacles to which solute can adhere, allowing for slow variation of obstacle size while requiring uniform cell size.

Here, we study the impact of slowly varying pore structure on macroscopic flow, transport, and sorption within a porous medium. Specifically, we consider steady flow through a heterogeneous, two-dimensional porous material comprising an array of solid obstacles. We allow for slow but arbitrary longitudinal variations in the size of obstacles, as in \citet{dalwadi2015understanding,dalwadi2016multiscale}, and also in their spacing. We begin by developing a general model for homogenised flow and transport for arbitrary obstacle shape, size, and spacing. We then develop detailed results for the simple case of circular obstacles.A key novelty of this approach is that allowing for two degrees of microstructural freedom affords a rich parameter space for exploration, including, for example, the ability to have a heterogeneous microstructure while maintaining uniform porosity, and allowing for a specific study of anisotropy. Mathematically, varying the longitudinal spacing requires dealing with a varying cell size in the homogenisation procedure. This is nontrivial, and adds a frequency modulation to the problem, as well as the typical amplitude modulation associated with homogenisation via the MMS \citep{chapman2011unified}.

For the flow, we assume steady Stokes flow with no-slip and no-penetration conditions on the solid surfaces. For solute transport, we consider transient advection and diffusion with removal via adsorption on the solid surfaces~(\S\ref{s:rad}). Following \citet{chapman2011unified, richardson2011derivation, bruna2015diffusion, dalwadi2015understanding, dalwadi2016multiscale}, we exploit the local periodicity of the pore geometry to homogenise the pore-scale problem via the MMS (\S\ref{sec:homogenisation}). 
Since we consider a microstructure in which both the size and the spacing of the solid obstacles vary slowly along the length of the porous material, the total area of each cell also varies slowly. The homogenisation method provides effective macroscopic equations for fluid flow, solute transport and sorption that are uniformly valid throughout the heterogeneous porous medium. For any chosen geometry, the permeability and diffusivity tensors we derive for any particular porous medium microstructure may be determined numerically. In this manuscript, to demonstrate the general approach we further calculate these tensors for a specific filter geometry comprising an array of circular obstacles arranged on a rectangular lattice. These tensors are strongly anisotropic, highlighting the fact that porosity alone is an insufficient measure of the pore-structure (\S\ref{sec:effective_quantities}). We use the homogenised model to investigate the effects of heterogeneous pore structure in a simple one-dimensional steady-state filtration problem (\S\ref{sec:1Dfilter}). Finally, we discuss the merits and limitations of the model (\S\ref{sec:conclusion}).

%%%%%%%%%%%%%%%%%%%%%%%%%%%%%%%%%%%%%%%%%%%%%%%%%%%%%%
\section{Model problem}
\label{s:rad}
%%%%%%%%%%%%%%%%%%%%%%%%%%%%%%%%%%%%%%%%%%%%%%%%%%%%%%

\begin{figure}
 \centering
 \includegraphics[width=\textwidth]{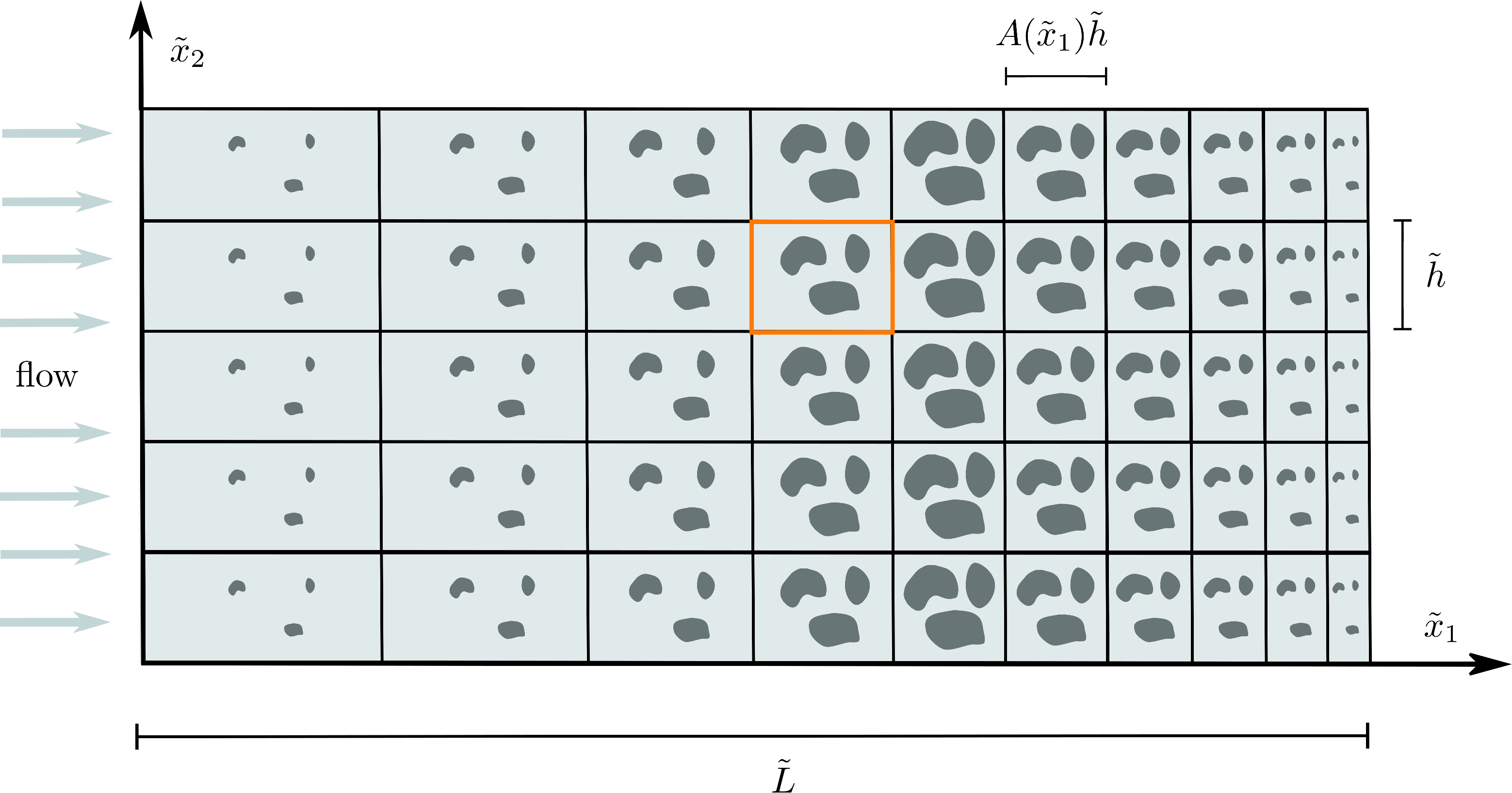} %Fig_Schem_dim_vary_R_arb_modified
 \caption{ \label{fig:schem_dim_1} We consider the flow of fluid carrying solute through a heterogeneous porous material in two dimensions.  The porous medium has length $\tilde{L}$ and is formed of an array of obstacles whose size depends only on a scale factor $\Lambda(\tilde{x}_1)$, located within each rectangular cell of transverse height $\tilde{h}$ and longitudinal width $A(\tilde{x}_1)\tilde{h}$. The porous medium is thus uniform in the transverse ($\tilde{x}_2$) direction but heterogeneous in the longitudinal ($\tilde{x}_1$) direction. We assume that the spacing between obstacles is small relative to the length of the porous medium: $\epsilon\defeq {\tilde{h}}/{\tilde{L}} \ll 1$. We isolate one cell in orange; this cell is shown in detail in Figure \ref{Fig_schem_2}. }
\end{figure}

We consider the steady flow of fluid carrying a passive solute through a rigid porous medium in two dimensions. The solute advects, diffuses and is removed via adsorption to the solid structure. The spatial coordinate is $\tilde{\bm{x}}\defeq\tilde{x}_1\bm{e}_1+\tilde{x}_2\bm{e}_2$, with $\tilde{x}_1$ and $\tilde{x}_2$ the dimensional longitudinal and transverse coordinates, respectively, and $\bm{e}_1$ and $\bm{e}_2$ the longitudinal and transverse unit vectors, respectively. The fluid enters the porous medium uniformly through the inlet at the left ($\tilde{x}_1=0$) and exits the porous medium through the outlet at the right ($\tilde{x}_1 = \tilde{L}$) (Figure~\ref{fig:schem_dim_1}). We denote dimensional quantities with a tilde.

The entire domain of the porous medium, denoted $\tilde{\Omega}$, comprises both the fluid and the solid structure of the domain. 
The latter constitutes an array of solid obstacles, as discussed in more detail below. We assume that the solute particles are small relative to the solid obstacles, and we measure the local density of solute (amount of solute per volume of fluid) via the concentration field $\tilde{c}(\tilde{\bm{x}},\tilde{t})$, where $\tilde{t}$ is dimensional time. This concentration field is defined within the fluid phase of the porous medium, denoted $\tilde{\Omega}_f$.

Note that we do not track solute once it has adsorbed to the solid surface, and we neglect any impact of this adsorption on the size of the obstacles. The latter point is justified by our assumption that the solute particles are negligible in size relative to the obstacles, and also because we are interested in macroscopic advective timescales, which are typically far shorter than those of solute accumulation and blocking.

The porous medium can be partitioned into an array of rectangular cells of fixed height $\tilde{h}$ and varying width $A(\tilde{x}_1)\tilde{h}$, where $A$ is the dimensionless aspect ratio. Each cell contains fixed and rigid obstacles of smooth but arbitrary shape. The shape of each obstacle is fixed and each obstacle can only grow or shrink isotropically about each obstacle's respective centre of mass according to a scale factor $\Lambda(\tilde{x}_1)$ --- that is, the obstacle size depends only on $\Lambda(\tilde{x}_1)$. The solid domain is the union of these obstacles, and is denoted $\tilde{\Omega}_s\defeq \tilde{\Omega}\setminus\tilde{\Omega}_f$. This construction leads to a porous medium whose properties vary in the longitudinal direction but not in the transverse direction (see Figure~\ref{fig:schem_dim_1}). We further assume that the porous medium is composed of a large number of obstacles in the longitudinal direction, which requires $\epsilon \defeq \tilde{h}/\tilde{L}\ll1$ with $A=\mathit{O}(1)$.

We assume that the fluid is incompressible and Newtonian, and that the flow is steady and dominated by viscosity. As such, the fluid velocity $\tilde{\bm{v}}(\tilde{\bm{x}})$ and pressure $\tilde{p}(\tilde{\bm{x}})$ satisfy the Stokes equations, subject to no-slip and no-penetration boundary conditions on the solid obstacles,
\begin{subequations}
\label{stokes_dim}
\begin{align}
\label{stokes_dim1}
-\tilde{\boldsymbol{\nabla}}\tilde{p}+\tilde{\mu}\tilde{\nabla}^2\tilde{\bm{v}} &= \bm{0}, \quad \tilde{\bm{x}}\in\tilde{\Omega}_f,\\
\tilde{\boldsymbol{\nabla}}\cdot\tilde{\bm{v}} &= 0, \quad \tilde{\bm{x}}\in\tilde{\Omega}_f,\\
\label{c}
\tilde{\bm{v}} &= \bm{0}, \quad \tilde{\bm{x}}\in \partial\tilde{\Omega}_s, 
\end{align}
\end{subequations}
where $\tilde{\mu}$ is the dynamic viscosity of the fluid, $\partial\tilde{\Omega}_s$ denotes the fluid--solid interface and $\tilde{\boldsymbol{\nabla}}$ is the gradient operator with respect to $\tilde{\bm{x}}$. 

We model solute transport and adsorption via the standard advection--diffusion equation with a linear, partially adsorbing condition at the fluid--solid interface:
\begin{subequations}\label{ad_diff_dim}
\begin{equation}
\frac{\partial \tilde{c}}{\partial \tilde{t}} = \tilde{\boldsymbol{\nabla}}\cdot\left(\tilde{\mathcal{D}}\tilde{\boldsymbol{\nabla}}\tilde{c}-\tilde{\bm{v}}\tilde{c}\right), \quad \tilde{\bm{x}}\in\tilde{\Omega}_f,
\end{equation}
\begin{equation}
\label{BC_orig}
-\tilde{\gamma}\tilde{c} = \tilde{\bm{n}}_s\cdot\left(\tilde{\mathcal{D}}\tilde{\boldsymbol{\nabla}}\tilde{c}-\tilde{\bm{v}}\tilde{c}\right), \quad \tilde{\bm{x}}\in \partial\tilde{\Omega}_s, 
\end{equation}
where $\tilde{\mathcal{D}}$ is the coefficient of molecular diffusion, $\tilde{\bm{n}}_s$ is the outward-facing unit normal to $\partial\tilde{\Omega}_s$, and $\tilde{\gamma} \geq 0$ 
is the constant adsorption coefficient.
Note that the second term on the right-hand side of Equation (\ref{BC_orig}) vanishes due to Equation (\ref{c}). 
Further, note that $\tilde{\gamma} = 0$ corresponds to no adsorption and $\tilde{\gamma}\to\infty$ corresponds to instantaneous adsorption, where the latter is equivalent to imposing $\tilde{c} = 0$ on $\partial\tilde{\Omega}_s$.
\end{subequations}

We define a function $\tilde{f}_s(\tilde{\bm{x}})$ such that on the fluid--solid interface $\partial \tilde{\Omega}_s$
\begin{equation}
\tilde{f}_s(\tilde{\bm{x}}) = 0. 
\end{equation}
We also define $\tilde{f}_s(\tilde{\bm{x}})>0$ inside the solid phase. Then, 
\begin{equation}
\label{nhat_dim}
\tilde{\bm{n}}_s(\tilde{\bm{x}}) \defeq \frac{\tilde{\bm{\nabla}}\tilde{f}_s}{\left|\tilde{\bm{\nabla}}\tilde{f}_s\right|},
\end{equation}
is the outward-facing normal to the fluid domain.

We make Equations (\ref{stokes_dim})--(\ref{ad_diff_dim}) dimensionless via the scalings
\begin{equation}\label{eq:non-dimensionalisation}
 \tilde{\bm{x}} = \tilde{L}\hat{\bm{x}}, \quad \tilde{\bm{v}} = \tilde{\mathcal{V}}\hat{\bm{v}}, \quad \tilde{p} = \left(\frac{\tilde{\mu} \tilde{\mathcal{V}}}{\epsilon^2\tilde{L}}\right)\hat{p}, 
\quad \tilde{c} = \tilde{\mathcal{C}} \hat{c}, \quad\text{and}\quad \tilde{t} = \left(\frac{\tilde{L}^2}{\tilde{\mathcal{D}}}\right)t,
\end{equation}
where $\tilde{{\mathcal{V}}}$ and $\tilde{\mathcal{{C}}}$ are the average inlet velocity and the average inlet concentration, respectively; $\bm{\hat{x}}$ and $t$ denote the dimensionless spatial and temporal coordinates, respectively; and $\hat{\bm{v}}= \hat{\bm{v}}(\bm{\hat{x}})$, $\hat{p}= \hat{p}(\bm{\hat{x}})$ and $\hat{c}= \hat{c}(\bm{\hat{x}},t)$ denote the dimensionless velocity, pressure and concentrations fields, respectively. This pressure scale balances the macroscopic pressure gradient against viscous dissipation at the pore-scale, as is standard in lubrication problems. Employing the scalings in Equation~(\ref{eq:non-dimensionalisation}), the flow problem (Eqs.~\ref{stokes_dim}) becomes 
\begin{subequations}\label{stokes}
 \begin{align}
 \label{Stoke_a}
 -\hat{\boldsymbol{\nabla}}\hat{p}+\epsilon^2 \hat{{\nabla}}^2{\hat{\bm{v}}} = \bm{0}, &\quad \hat{\bm{x}}\in\hat{\Omega}_f,\\
 \hat{\boldsymbol{\nabla}}\cdot{\hat{\bm{v}}} = 0, &\quad \hat{\bm{x}}\in\hat{\Omega}_f,\\
 \hat{\bm{v}} = \bm{0}, &\quad \hat{\bm{x}}\in \partial\hat{\Omega}_s,
 \end{align}
\end{subequations}
where $\hat{\bm{\nabla}}$ is the gradient operator with respect to $\hat{\bm{x}}$. 
Similarly, the transport problem (Eqs.~\ref{ad_diff_dim}) becomes 
\begin{subequations}\label{ad_diff}
 \begin{align}
 \label{c_hat}
 \frac{\partial {\hat{c}}}{\partial {t}} &= {\hat{\boldsymbol{\nabla}}}\cdot\left(\hat{{\boldsymbol{\nabla}}}{\hat{c}}-\Pen\ {\hat{\bm{v}}}\hat{c}\right), \ \quad \hat{\bm{x}}\in\hat{\Omega}_f,\\
 \label{Robin}
 -\epsilon\gamma\hat{c} &= \hat{\bm{n}}_s\cdot\left(\hat{{\boldsymbol{\nabla}}}\hat{c}-\Pen\ {\hat{\bm{v}}}\hat{c}\right), \quad \hat{\bm{x}}\in \partial\hat{\Omega}_s, 
 \end{align}
\end{subequations}
where $\hat{\bm{n}}_s(\hat{\bm{x}})$, a function of $\hat{\bm{x}}$, is the outward-facing normal to $\hat{\Omega}_f$ and the P\'{e}clet number $\Pen\defeq {\tilde{L}\tilde{\mathcal{V}}}/\tilde{\mathcal{D}}$ measures the rate of advective transport relative to that of diffusive transport and the dimensionless adsorption rate $\gamma\defeq{\tilde{\gamma} \tilde{L}}/{(\epsilon \tilde{\mathcal{D}})}$ measures the rate of adsorption relative to that of diffusive transport. Note that $\gamma\equiv {Da}/\epsilon$, where ${Da}=\tilde{\gamma}/\tilde{\mathcal{V}}$ is the Damk\"ohler number of the second kind. As discussed in more detail below, the subsequent analysis requires that $\Pen,\gamma = \mathit{O}(1)$ are constants independent of $\epsilon$; this represents a distinguished limit as highlighted below.

Finally, the dimensionless fluid--solid interface becomes $\hat{f}_s(\hat{\bm{x}}) = 0$ and Equation (\ref{nhat_dim}) becomes 
\begin{equation}
\label{nhat}
 \hat{\bm{n}}_s(\hat{\bm{x}}) \defeq \frac{\hat{\bm{\nabla}}\hat{f}_s}{\left|\hat{\bm{\nabla}}\hat{f}_s\right|}. 
\end{equation}

\section{Homogenisation}\label{sec:homogenisation}

Here, we approach the problem with homogenisation by the method of multiple scales (MMS). Classically, homogenisation via the MMS is an asymptotic technique for domains that can be represented as the union of a large number of strictly periodic cells \citep{chapman2008multiscale}; however, here we use an extension of the method to deal with materials with a locally periodic microstructure that can vary over the macroscale \citep{chapman2011unified, bruna2015diffusion,richardson2011derivation, dalwadi2015understanding}. The specific problem of circular obstacles slowly varying in size across the length of a filter has been considered in \citet{dalwadi2015understanding,dalwadi2016multiscale}, but this is a one-parameter variation in microstructure with the periodic cell size constant. In this paper, we generalise this approach to allow for arbitrary obstacle shape and include an additional degree of microstructural freedom in the spacing between obstacles. The latter extension allows us to explore porous media with novel properties such as a spatially varying microstructure but a spatially uniform porosity. In order to consider varying cell sizes, we must choose our microscale variable carefully to ensure microscale periodicity, in a similar manner to \citet{chapman2011unified,richardson2011derivation}.

Following the MMS, we isolate and solve the problem of flow and solute transport in an individual cell which is uniquely characterised by its aspect ratio 
\begin{equation}
 a(\hat{x}_1)=A(\tilde{x}_1) 
\end{equation}
and scale factor 
\begin{equation}
 \lambda(\hat{x}_1)= \Lambda(\tilde{x}_1). 
\end{equation}
We then construct a model for macroscopic flow and transport through the entire porous medium from the solution to these individual cell problems via local averaging. The result is a system of equations that are uniformly valid for all $\hat{\bm{x}}\in\hat{\Omega}$.

\subsection{Two spatial scales}

\begin{figure}
\centering
\includegraphics[width=\textwidth]{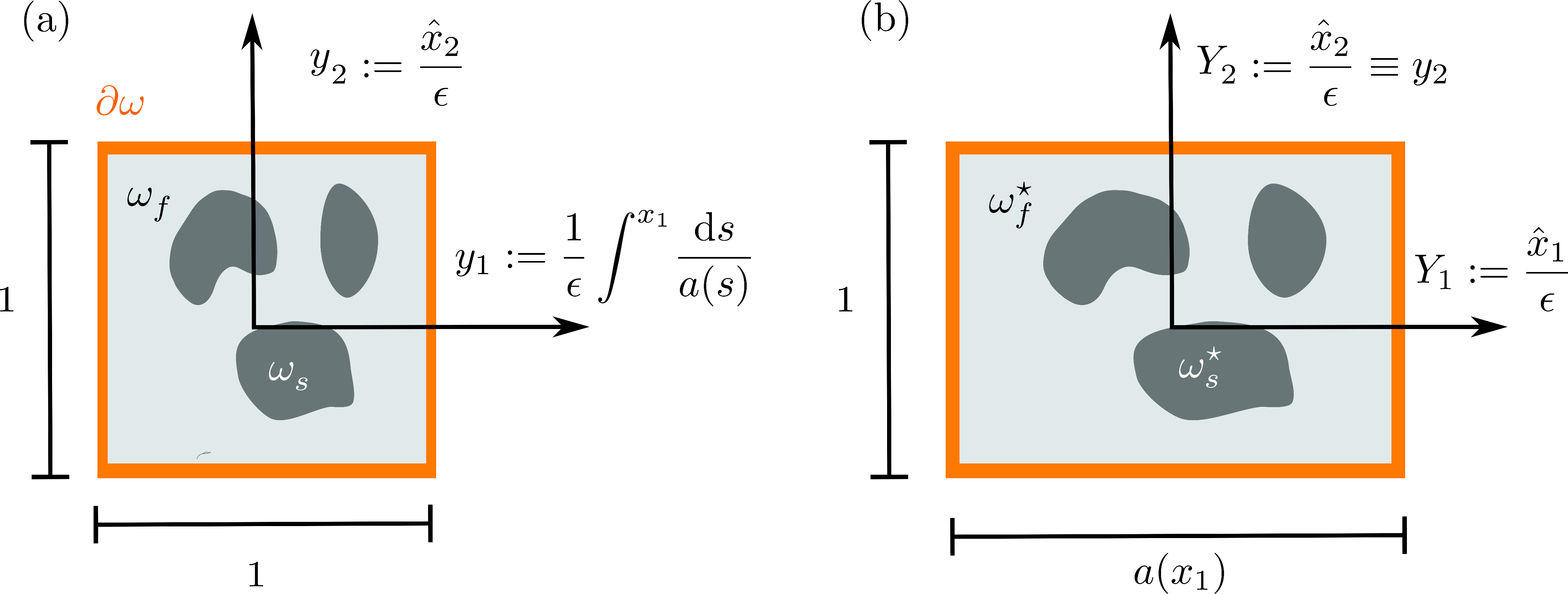} %Fig_Schem_2_y_and_Y_no_alpha_56.eps
\caption{An arbitrary cell within the porous medium (orange rectangle in Figure~\ref{fig:schem_dim_1}) represented in (a)~transformed microscale coordinates and (b)~naive microscale coordinates. The transformed microscale coordinates $y_1$ and $y_2$ and the naive microscale coordinates $Y_1$ and $Y_2$ are related via Equations~\eqref{y_map} and \eqref{Y_map}. The transformed microscale coordinates allow the slow variation in cell width $a$ to be scaled out of the cell problem, such that each naive rectangular cell is transformed into a square. \label{Fig_schem_2}}
\end{figure}

Applying the MMS as in \citet{chapman2011unified,richardson2011derivation}, we consider the spatial domain on two distinct length scales: the macroscale $\bm{x}\defeq \hat{\bm{x}}$, relative to which the the porous medium is of unit length, and a microscale coordinate $\bm{y}$ in which we are able to impose strict periodicity. This can be achieved via a mapping that transforms each cell (comprising the porous material) to a tessellating periodic cell. Here, we choose to transform to a square cell of unit area (see Figure \ref{Fig_schem_2}(a)). As such, our mapping will stretch the obstacles comprising the macroscale filter by a factor of $1/(\epsilon a(x_1))$ in the longitudinal direction and by $1/\epsilon$ in the transverse direction, \textit{i.e.,} 
\begin{equation}
\frac{\mathrm{d}y_1}{\mathrm{d}x_1} = \frac{1}{\epsilon a(x_1)} \quad \text{and} \quad \frac{\mathrm{d}y_2}{\mathrm{d}x_2} = \frac{1}{\epsilon},
\end{equation} 
which motivates the transformed microscale, defined by
\begin{equation}\label{y_map}
 {y_1}\defeq \frac{1}{\epsilon} \int^{x_1} \! \frac{\mathrm{d}s}{a(s)} \quad \text{and} \quad y_2 \defeq \frac{x_2}{\epsilon}. 
\end{equation}

Note that any arbitrary distribution of obstacles in the longitudinal directions (i.e., arbitrary longitudinal heterogeneity) can be imposed by fixing the functions $a(x_1)$ and $\lambda(x_1)$, while in the transverse direction the porous medium is exactly periodic. We note that additional heterogeneity in the transverse direction can be considered through a more general mapping \citep{richardson2011derivation}.

To understand the implications of the mapping (Eq.~\ref{y_map}) on a single cell, we note that (on the macroscale) the domain of the porous medium, $\bm{x}\in\Omega$, comprises a fluid domain $\Omega_f$ and a complementary solid domain $\Omega_s$. A single cell can be obtained by discretising the porous medium intorectangular cells of height $\epsilon$ and width $\epsilon a(x_1)$.For any single rectangular cell in the domain, the transformation (Eq.~\ref{y_map}) yields a transformed microscale, $\bm{y}$. On the transformed microscale, each cell $\omega$ comprises a fluid phase $\omega_f(x_1)$ and solid obstacles, the union of which is denoted $\omega_s(x_1)\defeq \omega \setminus \omega_f(x_1)$. The fluid--solid interface $\partial\omega_s(x_1)$ is the union of the boundary of the obstacles. Each cell has four additional boundaries that separate it from neighbouring cells. We denote the top and bottom boundaries $\partial \omega_{=}$ and the left and right boundaries $\partial \omega_{||}$ with the union of these being the unit cell boundary $\partial\omega$. 

Following \citet{chapman2011unified, richardson2011derivation}, we choose the microscale mapping such that the the cell size is the same throughout the domain. Hence, for example, circles will approximately map to ellipses. Since the untransformed cell size varies spatially through the domain, this microscale mapping will lead to obstacles that vary by an $\mathit{O}(\epsilon)$ amount between neighbouring cells, but by an $\mathit{O}(1)$ amount over the macroscale. We systematically account for these variations using the methodology presented in \emph{e.g.} \citet{bruna2015diffusion,dalwadi2015understanding}.

Finally, we note that the transformed microscale variable can be difficult to interpret physically. As such, it will be helpful to define a `naive' microscale coordinate \begin{equation}
\label{Y_map}
 \bm{Y}\defeq\bm{x}/\epsilon,
\end{equation}
in which each cell is of unit transverse height but of longitudinal width $a(x_1)$ (Figure \ref{Fig_schem_2}(b)). After completing the homogenisation procedure in the transformed microscale (Eq.~\ref{y_map}), we will transform the relevant cell problems to the naive microscale (Eq.~\ref{Y_map}), in order to present them more intuitively and subsequently solve them numerically. Note that the domains and boundaries in the (naive) rectangular $\bm{Y}$-cell will be denoted as in the square $\bm{y}$-cell, but with the addition of a superscript~$\star$. Further, we emphasise that the microscale (cell) problems we derive and solve are not physical flow or transport problems, but rather mathematical constructs that enable us to invoke the MMS.

We now perform the homogenisation. Following the MMS, we take $\bm{x}$ and $\bm{y}$ to be independent spatial parameters. We therefore rewrite all functions of $\hat{\bm{x}}$ as functions of $\bm{x}$ and $\bm{y}$: $\hat{\bm{v}}(\hat{\bm{x}})\defeq\bm{v}(\bm{x},\bm{y})$, $\hat{p}(\hat{\bm{x}}) \defeq p(\bm{x},\bm{y})$, and $\hat{c}(\hat{\bm{x}},t) \defeq c(\bm{x},\bm{y},t)$. Note that for functions that are dependant of $\bm{Y}$ in lieu of $\bm{y}$ we adorn the respective function with a superscript $\star$. Spatial derivatives then become 
\begin{subequations}
\label{spatial_map}
\begin{align}
\frac{\partial}{\partial \hat{x}_i} = \frac{\partial}{\partial x_i} + \frac{\sigma_{ij}}{\epsilon} \frac{\partial}{\partial y_j},
\end{align}
for $i,j = 1, 2$, and where $\sigma_{ij}=\left(\bm{\sigma}\right)_{ij}$ and 
\begin{equation}
 \bm{\sigma}=\begin{pmatrix}
 \displaystyle\frac{1}{a(x_1)} & 0\\
 0 & 1
\end{pmatrix}.
\end{equation}
Alternatively, in vector form, the spatial derivatives become
\begin{align}
\hat{\bm{\nabla}} \defeq \bm{\nabla}_x+\frac{1}{\epsilon}\bm{\nabla}_y^a
\end{align}
where $\boldsymbol{\nabla}_x$ is the gradient operator with respect to the coordinate $\bm{x}$ and where
\begin{equation}
\bm{\nabla}_y^a\defeq \left(\frac{1}{a}\frac{\partial}{\partial y_1},\frac{\partial}{\partial y_2}\right)^\intercal,
\end{equation}
is the gradient operator associated with the $\bm{y}$-coordinate transform. 
\end{subequations}
For a given quantity $Z(\bm{x},\bm{y},t) = Z^\star(\bm{x},\bm{Y},t)$ , there are two different averages of interest: the intrinsic (fluid) average
\begin{equation}
 \langle Z\rangle(\bm{x},t) \defeq \frac{1}{|\omega_f(x_1)|} \! \int_{\omega_f(x_1)} Z(\bm{x},\bm{y},t) \, \mathrm{d}S_y \equiv \frac{1}{|\omega_f^{\star} (x_1)|}\! \int_{\omega_f^\star (x_1)} Z^\star(\bm{x},\bm{Y},t) \, \mathrm{d}S_Y,
\end{equation}
where the total fluid area in the transformed cell $|\omega_f|$ (or naive cell $|\omega_f^{\star}|$) is a function of $a(x_1)$ and $\lambda(x_1)$;
and the volumetric average
\begin{equation}
\label{volume_int_y2Y}
 \frac{1}{|\omega(x_1)|}\int_{\omega(x_1)} \! Z(\bm{x},\bm{y},t)\, \mathrm{d} S_y \equiv \frac{1}{|\omega^{\star}(x_1)|}\int_{\omega^{\star}(x_1)} \! Z^\star(\bm{x},\bm{Y},t) \, \mathrm{d}S_Y, 
\end{equation}
where $|\omega|=1$ and $|\omega^{\star}|=a$. Here, $\mathrm{d}S_y\defeq\mathrm{d}y_1\mathrm{d}y_2$ is an area element of the transformed microscale fluid region, $\mathrm{d}S_Y\defeq\mathrm{d}Y_1\mathrm{d}Y_2$ is an area element of the naive microscale fluid region and the porosity $\phi$ is 
\begin{equation}
\label{phi}
\phi(x_1) = \frac{|\omega_f(x_1)|}{|\omega(x_1)|}\equiv |\omega_f(x_1)| \left(= \frac{|\omega_f^{\star}(x_1)|}{|\omega^{\star}(x_1)|} \right).
\end{equation}
Thus, $\langle c\rangle$ is the amount of solute per unit fluid area within the porous medium, while $\phi\langle c\rangle$, the volumetric average of the concentration, is the amount of solute per unit total area. 

We define the average velocity, pressure, and concentration as
\begin{equation}
\label{C}
 \bm{V}(\hat{\bm{x}}) \equiv \bm{V}(\bm{x}) \defeq \langle \bm{v} \rangle , \quad P(\hat{\bm{x}}) \equiv P(\bm{x}) \defeq \langle p \rangle, \quad \text{and} \quad C(\hat{\bm{x}},t) \equiv C(\bm{x},t) \defeq \langle c \rangle,
\end{equation}
respectively.Note that $\phi \bm{V} $ is the standard Darcy flux.

\subsection{Flow problem}
\label{sec: Flow problem}

For a passive tracer as described in this paper, the flow problem (Eqs.~\ref{Stoke_a}) does not depend on~$c$. Using Equation~(\ref{spatial_map}), Equations~(\ref{stokes}) in an arbitrary cell become 

\begin{subequations}
\label{stokes_homog}
\begin{equation}
\label{stokes_main}
-\left(\boldsymbol{\nabla}_x+\frac{1}{\epsilon}\boldsymbol{\nabla}_y^a\right){p}+\left(\epsilon^2\nabla^2_x+\epsilon\boldsymbol{\nabla}_x\cdot\boldsymbol{\nabla}_y^a +\epsilon\boldsymbol{\nabla}_y^a\cdot\boldsymbol{\nabla}_x +(\nabla_y^a)^2\right)\bm{v} = \bm{0}, \quad {\bm{y}}\in\omega_f(x_1),
\end{equation}
\vspace{-4mm}
\begin{align}
\label{incomp}
\left(\epsilon\boldsymbol{\nabla}_x+\boldsymbol{\nabla}_y^a\right)\cdot{{\bm{v}}} &= 0, \quad {\bm{y}}\in\omega_f(x_1) \\
\label{no_slip_orig}
\bm{v} &= \bm{0}, \quad {\bm{y}}\in \partial\omega_s(x_1).
\end{align}
For clarity of presentation in what follows, we have multiplied by $\epsilon$ during the derivation of  Equation (\ref{incomp}).

To proceed using the MMS, we must also impose periodicity of $\bm{v}$, $p$ and $c$ over a single microscale cell (\textit{i.e.,} local periodicity). Enforcing periodicity of all quantities at both the top and bottom, $\partial\omega_{=}$, and left and right, $\partial\omega_{||}$, cell boundaries leads to
\begin{equation}
\bm{v} \quad \text{and}\quad p \quad \text{periodic on} \quad \bm{y}\in\partial\omega_{=}\ \ \text{and}\ \ \partial\omega_{||}. 
\end{equation}
\end{subequations}

We now seek an asymptotic solution to Equations (\ref{stokes_homog}) by expanding $\bm{v}$ and $p$ in powers of~$\epsilon$: 
\begin{subequations}
\begin{align}
& \bm{v}(\bm{x},\bm{y}) = \bm{v}^{(0)}(\bm{x},\bm{y})+\epsilon\bm{v}^{(1)}(\bm{x},\bm{y})+\cdots \quad \text{as}\quad \epsilon\to0, \\
& p(\bm{x},\bm{y} ) = p^{(0)}(\bm{x},\bm{y})+\epsilon p^{(1)}(\bm{x},\bm{y})+\cdots \quad \ \text{as}\quad \epsilon\to0.
\end{align}
\end{subequations}
Considering terms of $\mathit{O}\left(1/\epsilon\right)$ in Equations (\ref{stokes_main}) gives
\begin{equation}\label{Operator}
 \boldsymbol{\nabla}_y^ap^{(0)}= \bm{0}, 
\end{equation}
from which we conclude the standard result that, at leading order, the pressure is uniform on the microscale: $p^{(0)} = p^{(0)}(\bm{x})$.

Considering terms of $\mathit{O}\left(1\right)$ in Equation (\ref{stokes_homog}) gives 
\begin{subequations}\label{stokes_homog_O_ep}
 \begin{align}
 \label{stokes_main_O_ep}
-\boldsymbol{\nabla}_x p^{(0)}-\boldsymbol{\nabla}_y^a{p^{(1)}}+(\nabla_y^a)^2\bm{v}^{(0)} = \bm{0}, &\quad {\bm{y}}\in\omega_f(x_1),\\
\label{incomp_O_ep}
\boldsymbol{\nabla}_y^a\cdot{{\bm{v}^{(0)}}} = 0, &\quad {\bm{y}}\in\omega_f(x_1),\\
\label{no_slip_no_pen}
{\bm{v}^{(0)}} = \bm{0}, &\quad {\bm{y}}\in \partial\omega_s(x_1), 
\end{align}
with
\begin{equation}
\bm{v}^{(0)}\quad \text{and}\quad p^{(1)} \quad \text{periodic on} \quad \bm{y}\in\partial\omega_{=}\ \ \text{and}\ \ \partial\omega_{||}.
\end{equation}
\end{subequations}
The form of Equations (\ref{stokes_homog_O_ep}) suggest that we can scale $\bm{\nabla}_x p^{(0)}$ out of the problem via the substitutions
\begin{subequations}
\label{v0_p1}
\begin{align}
\label{Darcy_micro}
\bm{v}^{(0)} &= -\bm{{\mathcal{K}}}(\bm{x}, \bm{y}) \cdot
\boldsymbol{\nabla}_x p^{(0)},\\
p^{(1)} &= -\bm{\Pi}(\bm{x},\bm{y})\cdot\boldsymbol{\nabla}_x p^{(0)} +\breve{p}(\bm{x}),
\end{align}
\end{subequations}
where $\breve{p}(\bm{x})$ is a scalar function, $\bm{{\mathcal{K}}}(\bm{x}, \bm{y})$ is a tensor function, and $\bm{\Pi}(\bm{x},\bm{y})$ is a vector function. 
Using Equations (\ref{v0_p1}) and the fact that $p^{(0)}$ is independent of $\bm{y}$, Equations~\eqref{stokes_homog_O_ep} become
\begin{subequations}\label{stokes_homog_O_ep_scaled_1}
 \begin{align}
 \left(\bm{I}-\boldsymbol{\nabla}_y^a\otimes\bm{\Pi}+(\nabla^a_y)^2\bm{{\mathcal{K}}}\right)\cdot\boldsymbol{\nabla}_x p^{(0)} = \bm{0}, &\quad {\bm{y}}\in\omega_f(x_1), \\
 \left(\boldsymbol{\nabla}_y^a\cdot{\bm{{\mathcal{K}}}}\right)\cdot\boldsymbol{\nabla}_x p^{(0)} = \bm{0}, &\quad {\bm{y}}\in\omega_f(x_1), \\
 \bm{{\mathcal{K}}}\cdot\boldsymbol{\nabla}_x p^{(0)} = \bm{0}, &\quad {\bm{y}}\in \partial\omega_s(x_1), 
\end{align}
with
\begin{equation}
 \mathcal{K}_{ij}\defeq\left(\bm{\mathcal{K}}\right)_{ij} \ \ \text{and} \ \ \Pi_{i}\defeq\left(\bm{\Pi}\right)_{i} \quad \text{periodic on} \quad \bm{y}\in\partial\omega_{=} \ \ \text{and}\ \ \partial\omega_{||},
\end{equation}
where $\bm{I}$ is the identity tensor and where
\begin{equation} \left(\boldsymbol{\nabla}_y^a\otimes\bm{\Pi}\right)_{ij} = \sigma_{ik} \frac{\partial \Pi_j}{\partial y_k} \quad \text{and} \quad(\boldsymbol{\nabla}_y^a\cdot{\bm{{\mathcal{K}}}})_{i} = \sigma_{jk} \frac{\partial {\mathcal{K}}_{ji}}{\partial y_k}.
\end{equation}
\end{subequations}
Note that in the above we have adopted the summation convention; we will adopt the summation convention throughout this manuscript.
Equations (\ref{stokes_homog_O_ep_scaled_1}) must hold for arbitrary $\boldsymbol{\nabla}_x p^{(0)}$, hence $\bm{{\mathcal{K}}}(\bm{x}, \bm{y})$ and $\bm{\Pi}(\bm{x},\bm{y})$ must satisfy the system
\begin{subequations}
\label{stokes_homog_O_ep_scaled}
 \begin{align}
 \bm{I}-\boldsymbol{\nabla}_y^a\otimes\bm{\Pi}+(\nabla^a_y)^2\bm{{\mathcal{K}}} = \bm{0}, &\quad {\bm{y}}\in\omega_f(x_1), \label{stokes_main_O_ep_scaled} \\
 \boldsymbol{\nabla}_y^a\cdot{\bm{{\mathcal{K}}}} = \bm{0}, &\quad {\bm{y}}\in\omega_f(x_1), \label{incomp_O_ep_scaled} \\
 \bm{{\mathcal{K}}} = \bm{0}, &\quad {\bm{y}}\in \partial\omega_s(x_1),
\end{align}
with
\begin{equation}
 {\mathcal{K}}_{ij} \ \ \text{and} \ \ \Pi_{i} \quad \text{periodic on} \quad \bm{y}\in\partial\omega_{=} \ \ \text{and}\ \ \partial\omega_{||}. 
\end{equation}
\end{subequations}

In general, Equations (\ref{stokes_homog_O_ep_scaled}) must be solved numerically for each desired cell geometry (\textit{i.e.,} pairs of $a$ and $\lambda$). Note that Equations (\ref{stokes_homog_O_ep_scaled}) are independent of $\bm{\nabla}_xp^{(0)}$, justifying our scalings in Equation~\eqref{v0_p1}.

To derive a macroscale relationship between velocity and pressure from Equation (\ref{Darcy_micro}), we expand the averaged quantities defined in Equation (\ref{C}) in powers of $\epsilon$:
\begin{subequations}
\begin{align}
\label{V_exp}
 \bm{V}(\hat{\bm{x}}) &= \bm{V}^{(0)}(\hat{\bm{x}})+\epsilon\bm{V}^{(1)}(\hat{\bm{x}})+\cdots\quad \text{as}\quad \epsilon\to0, \\
 P(\hat{\bm{x}}) &= P^{(0)}(\hat{\bm{x}})+\epsilon P^{(1)}(\hat{\bm{x}})+\cdots\quad\ \text{as}\quad \epsilon\to0.
\end{align}
\end{subequations}
Note that
 \begin{equation}
 P^{(0)}(\hat{\bm{x}}) = \langle p^{(0)}(\bm{x})\rangle \equiv p^{(0)}(\bm{x}),
\end{equation}
since $p^{(0)}$ is independent of $\bm{y}$. We then take the intrinsic average of Equation (\ref{Darcy_micro}) to determine that the leading-order macroscale velocity depends on gradients in the leading-order macroscale pressure according to Darcy's law:
\begin{subequations}
\label{solvability}
\begin{equation}
\label{Darcy}
 \phi\bm{V}^{(0)} = -\bm{K}(\phi, a)\cdot\hat{\boldsymbol{\nabla}}P^{(0)},
\end{equation}
where we have introduced the macroscale permeability tensor
\begin{equation}
\label{K_def}
\bm{K}(\phi, a) \defeq \phi\langle{\bm{\mathcal{K}}}\rangle,
\end{equation}
and where $\phi$ and $a$ are known functions of $\hat{{x}}_1$. Prescribing both $\phi$ and $a$ determines $\lambda$ via a simple geometric relation, specific to the chosen geometry of the porous material.
\end{subequations}

Being averaged in $\bm{y}$, Equation (\ref{Darcy}) depends on $\hat{\bm{x}} = \bm{x}$ only and we have therefore replaced $\bm{\nabla}_x$ with $\hat{\bm{\nabla}}$. If the cell geometry has symmetric reflectional symmetry along both the $y_1$ and $y_2$ axes, the symmetry of the boundary conditions imply that $\bm{K}$ is diagonal and further, if $a=1$ also, then $\bm{K}$ reduces to a scalar multiple of $\boldsymbol{I}$.

Equation (\ref{Darcy}) provides two equations for three unknowns. To develop another constraint in terms of $\bm{V}^{(0)}$ and $P^{(0)}$, consider the $\mathit{O}(\epsilon)$ terms from Equations (\ref{incomp}) and (\ref{no_slip_orig}):
\begin{subequations}
\begin{align}
\label{O1incomp}
 \bm{\nabla}_x\cdot\bm{v}^{(0)}=-\bm{\nabla}_y^a\cdot\bm{v}^{(1)},\quad &\bm{y}\in\omega_f\\
 \label{v10}
\bm{v}^{(1)} = \bm{0}, \quad &\bm{y}\in\partial\omega_s.
\end{align}
\end{subequations}
We take the intrinsic average of Equation (\ref{O1incomp}) and apply the divergence theorem to the right-hand side which vanishes by Equation (\ref{v10}). Then, applying the transport theorem (Eq.~\ref{transport_final}), derived in Appendix \ref{transport_thm}, to the left-hand side of the intrinsic average of Equation~\eqref{O1incomp} yields
\begin{equation}
\label{macro_incomp}
 \bm{\nabla}_x\cdot\int_{\omega_f}\bm{v}^{(0)}\mathrm{d}S_y = 0, 
\end{equation}
where we have used Equation~(\ref{no_slip_no_pen}). Expressing Equation~\eqref{macro_incomp} in terms of the averaged quantity $\bm{V}^{(0)}$ gives
\begin{equation}
\label{Incomp_macro}
 \hat{\boldsymbol{\nabla}}\cdot(\phi\bm{V}^{(0)}) = 0,
\end{equation} 
which closes the system defined in Equations (\ref{Darcy}). Similarly to Equation~\eqref{Darcy}, there is no $\bm{y}$-dependence in Equation (\ref{Incomp_macro}), so we have replaced $\bm{\nabla}_x$ with $\hat{\bm{\nabla}}$. 

To evaluate $\bm{K}$, we find it convenient to map the system (Eq.\ref{stokes_homog_O_ep_scaled}) to the naive microscale coordinate $\bm{Y}$, defined in Equation~\eqref{Y_map}. In the naive microscale coordinate, different values of $a$ manifest as physical changes to the domain rather than as changes to the governing equations, yielding more intuitive cell problems. This mapping gives
\begin{subequations}
\label{stokes_homog_O_ep_scaled_naive}
 \begin{align}
 \bm{I}-\boldsymbol{\nabla}_Y\otimes\bm{\Pi}^\star+\nabla_Y^2\bm{{\mathcal{K}}}^\star = \bm{0}, &\quad {\bm{Y}}\in\omega_f^{\star}(x_1), \label{stokes_main_O_ep_scaled_naive} \\
 \boldsymbol{\nabla}_Y\cdot{\bm{{\mathcal{K}}}}^\star = \bm{0}, &\quad {\bm{Y}}\in\omega_f^{\star}(x_1), \label{incomp_O_ep_scaled_naive} \\
 \bm{{\mathcal{K}}}^\star = \bm{0}, &\quad {\bm{Y}}\in \partial\omega_s^{\star}(x_1),
\end{align}
with 
\begin{equation}
 \mathcal{K}^\star_{ij}\defeq\left(\bm{\mathcal{K}}^\star\right)_{ij} \ \ \text{and} \ \ \Pi_{i}^\star\defeq\left(\bm{\Pi}^\star\right)_{i} \quad \text{periodic on} \quad \bm{Y}\in\partial\omega_{=}^{\star} \ \ \text{and}\ \ \partial\omega_{||}^{\star}, 
\end{equation} 
where $\boldsymbol{\nabla}_Y$ is the gradient operator with respect to the coordinate $\bm{Y}$. Note that
\begin{equation}
\label{eq: tensor def}\left(\boldsymbol{\nabla}_Y\otimes\bm{\Pi}^\star\right)_{ij} = \frac{\partial \Pi_j^\star}{\partial Y_i} \quad \text{and} \quad(\boldsymbol{\nabla}_Y\cdot{\bm{{\mathcal{K}}}}^\star)_{i} = \frac{\partial {\mathcal{K}}^\star_{ji}}{\partial Y_j}.
\end{equation}
\end{subequations}
In \S\ref{figs} we consider a porous medium with a simple, prescribed microstructure. In that section, we solve Equations~\eqref{stokes_homog_O_ep_scaled_naive} using COMSOL Multiphysics\textsuperscript{\textregistered}, graphically present $\bm{K}(\phi, a)$, and discuss its implications.
 
\subsection{Transport problem}

We now perform a similar homogenisation procedure for the solute-transport problem (Eqs. \ref{ad_diff}). The main difference between the classic homogenisation procedure and the homogenisation we carry out here is that we use the transformed microscale $\bm{y}$ to convert a locally periodic tessellating cell structure into a strictly periodic tessellating cell structure. As such, we proceed following the framework of \citet{chapman2011unified} and \citet{richardson2011derivation}. A key step is to consider the unit normal $\hat{\bm{n}}_s$ that appears in Equation (\ref{Robin}). In general, under the microscale transformation (Eq.~\ref{y_map}), $\hat{\bm{n}}_s$ will \emph{not} be transformed to the geometric normal of the transformed cell. Hence, we must take care when transforming the normal into multiple scales form.

Under the multiple scales framework, the unit normal to the solid interface is written as a function of both the macro- and microscales: $\hat{\bm{n}}_s(\hat{\bm{x}}) = \bm{n}_s(\bm{x},\bm{y})$, and similarly for the function $\hat{f}_s(\hat{\bm{x}}) = f_s(\bm{x},\bm{y})$, which vanishes on the solid interface. The consistent transformation of $\bm{n}_s \equiv n_s^{i} \bs{e}_i$ requires the consistent application of the MMS derivative transformation (Eq.~\ref{spatial_map}) to the definition of $\bm{n}_s$ in terms of $f_s$ given by Equation~(\ref{nhat}), to obtain the transformed unit normal
\begin{subequations}
\label{eq: normals} 
\begin{align}
\label{n} 
\bm{n}_s = \frac{\left(\boldsymbol{\nabla}_y^a + \epsilon \boldsymbol{\nabla}_x\right)f_s}{\left|\left(\boldsymbol{\nabla}_y^a + \epsilon \boldsymbol{\nabla}_x \right)f_s\right|} = \frac{\displaystyle \left(\sigma_{ij} \frac{\partial f_s}{\partial y_j} + \epsilon \frac{\partial f_s}{\partial x_i}\right) \bs{e}_i }{\left[\displaystyle \sigma_{kl}\sigma_{km} \frac{\partial f_s}{\partial y_l}\frac{\partial f_s}{\partial y_m} \right]^{1/2} + \mathit{O}(\epsilon)}. 
 \end{align}
It will also be helpful to define the leading-order transformed unit normal $\bm{n}^{Y} = n^Y_i \bs{e}_i$ as follows
\begin{align}
\label{eq: NY}
\bm{n}^{Y} = \frac{\displaystyle \sigma_{ij} \frac{\partial f_s}{\partial y_j} \bs{e}_i}{\left[\displaystyle \sigma_{kl}\sigma_{km} \frac{\partial f_s}{\partial y_l}\frac{\partial f_s}{\partial y_m} \right]^{1/2}},
\end{align}
such that $\bm{n}_s \sim \bm{n}^{Y}$ as $\eps \to 0$. However, we also note that the geometric unit normal $\bm{n}^y = n^y_i \bm{e}_i$ is defined as
\begin{align}
 \label{eq: n geo} 
 \bm{n}^y := \frac{\boldsymbol{\nabla}_y f_s}{\left|\boldsymbol{\nabla}_y f_s \right|} = \frac{\displaystyle \frac{\partial f_s}{\partial y_i} \bs{e}_i}{\left[\displaystyle \frac{\partial f_s}{\partial y_j} \frac{\partial f_s}{\partial y_j} \right]^{1/2}},
\end{align}
where $\boldsymbol{\nabla}_y$ is the gradient operator with respect to the coordinate $\bm{y}$. Importantly, the transformed normal (Eq.~\ref{n}) and geometric normal (Eq.~\ref{eq: n geo}) are not equal. Moreover, comparing Equation~\eqref{eq: NY} and Equation~\eqref{eq: n geo} reveals that they are not even equal to leading order in $\epsilon$ (unless $a \equiv 1$).

To facilitate our subsequent manipulation of the transformed problem, it will be helpful to write the transformed normal $\bm{n}_s$ in terms of the geometric normal $\bm{n}^y$. Since Equation~\eqref{eq: n geo} can be rearranged to obtain $\partial f_s /\partial y_i = \left|\boldsymbol{\nabla}_y f_s \right| n^y_i$, we can re-write the transformed normal (Eq.~\ref{n}) as
\begin{align}
\label{n transform in geo} 
 \bs{n}_s =
 \frac{\displaystyle \left(\sigma_{ij} n^y_j + \epsilon N_i\right) \bs{e}_i}{\left[\displaystyle \sigma_{kl}\sigma_{km} n^y_ln^y_m \right]^{1/2} + \mathit{O}(\epsilon)},
\end{align}
where the macroscale perturbation to the normal $\bm{N} = N_i \bm{e}_i$ is defined as
\begin{equation}\label{normal_N}
 \bm{N}\defeq\frac{\bm{\nabla}_xf_s}{|\bm{\nabla}_y f_s|}.
\end{equation}
The macroscale perturbation to the normal $\bm{N}$ formally quantifies the effect of the transformed microscale structure varying over the macroscale within the MMS framework.
\end{subequations}

Having defined the transformed normal in terms of the geometric normal, we are now in a position to proceed with the homogenisation. Under the spatial transformations (Eq.~\ref{spatial_map}), Equations (\ref{ad_diff}) become 
\begin{subequations}\label{ad_diff_scaled}
\begin{equation}\label{c_XY}
 \epsilon\frac{\partial {{c}}}{\partial {t}} = \left(\epsilon \frac{\partial}{\partial x_i}+\sigma_{ij} \frac{\partial}{\partial y_j}\right) \left[\frac{\partial c}{\partial x_i}+\dfrac{\sigma_{ik}}{\epsilon}\frac{\partial c}{\partial y_k}- \Pen \ {v_i}{c}\right], \quad {{\bm{y}}}\in\omega_f(x_1),
\end{equation}
\begin{equation}\label{Robin_2_scales}
-\epsilon\gamma{c} \left[\displaystyle \sigma_{kl}\sigma_{km} n^y_ln^y_m \right]^{1/2} +\mathit{O}(\epsilon^2)= \left(\sigma_{ij} n^y_j + \epsilon N_i\right) \left[\frac{\partial c}{\partial x_i}+\dfrac{\sigma_{ik}}{\epsilon}\frac{\partial c}{\partial y_k}- \Pen \ {v_i}{c}\right], \quad {{\bm{y}}}\in \partial\omega_s(x_1), 
\end{equation}
with
\begin{equation}
 v_i, \quad c, \quad \text{periodic on} \quad \bm{y}\in\partial\omega_{=}\ \ \text{and}\ \ \partial\omega_{||},
\end{equation}
\end{subequations}
where $\bm{v} = v_i \bm{e}_i$. Note that, for clarity of presentation in what follows, we have multiplied by $\epsilon$ when deriving Equation (\ref{c_XY}) from Equation~\eqref{c_hat}. We now consider an expansion of the concentration field of the form 
\beq
\label{eq:c_expansion}
c(\bm{x}, \bm{y},t) = c^{(0)}(\bm{x}, \bm{y},t) + \eps c^{(1)}(\bm{x}, \bm{y},t) + \eps^2 c^{(2)}(\bm{x}, \bm{y},t) +\cdots \quad \mbox{as} \quad \eps \to 0. 
\eeq
Note that we take $\Pen, \gamma = \mathit{O}(1)$ to be constants independent of $\epsilon$; this corresponds to a distinguished limit where all the transport mechanisms balance over the macroscale (\textit{cf.,} Equation~(\ref{eq:c_homogenised final})).
Considering Equation~\eqref{ad_diff_scaled} at leading order --- that is, $\mathit{O}(1/\epsilon)$ --- we obtain
\begin{subequations}
\label{eq:c_zeroorder}
\begin{align}
 \sigma_{ij} \sigma_{ik} \dfrac{\partial^2 c^{(0)}}{\partial y_j\partial y_k} = 0, \qquad &\bm{y} \in \omf(x_1), \\ 
 \sigma_{ij} \sigma_{ik} n_j^y \dfrac{\partial c^{(0)}}{\partial y_k} = 0, \qquad &\bm{y} \in \pa \oms({x_1}), \\ 
 \quad{v}^{(0)}_i, \quad c^{(0)}\quad \text{ periodic on} \quad &\bm{y}\in\partial\omega_{=}\ \ \text{and}\ \ \partial\omega_{||}.
\end{align}
\end{subequations}
By inspection, we find that $c^{(0)} = c^{(0)}(\bm{x},t)$ is a nontrivial solution to this system. By linearity, this solution is unique and therefore the leading-order concentration is independent of $\bm{y}$.

Considering Equations (\ref{ad_diff_scaled}) at $\mathit{O}(1)$, we obtain 
\begin{subequations}
\label{eq:c_1storder}
 \begin{align}
\label{Laplace_c1}
 \sigma_{ij} \sigma_{ik} \dfrac{\partial^2 c^{(1)}}{\partial y_j\partial y_k} = 0, \qquad &\bm{y} \in \omf(x_1), \\
\label{BC_c1}
 \sigma_{ij} \sigma_{ik} n_j^y \dfrac{\partial c^{(1)}}{\partial y_k} = -\sigma_i n_i^y \dfrac{\partial c^{(0)}}{\partial x_i}, \qquad &\bm{y} \in \pa \oms({x_1}), \\
 \label{BC_c2}
 \quad{v}^{(1)}_i, \quad c^{(1)} \quad \text{periodic on} \quad &\bm{y}\in\partial\omega_{=}\ \ \text{and}\ \ \partial\omega_{||}, 
\end{align}
\end{subequations}
where we have used $c^{(0)} = c^{(0)}(\bm{x},t)$, microscale incompressibility (Eq.~\ref{incomp_O_ep}), and the no-slip and no-penetration conditions (Eq.~\ref{no_slip_no_pen}) on the solid surface. The form of Equations~\eqref{eq:c_1storder} suggest that we can scale $\bm{\nabla}_x c^{(0)}$ out of the problem via the substitution
\begin{align}
\label{eq:Gamma}
c^{(1)}(\bm{x}, \bm{y},t) = - \dfrac{\partial c^{(0)}}{\partial x_l} \Gamma_n (\bm{x}, \bm{y}) + \breve{c}(\bm{x},t),
\end{align}
where $\breve{c}$ is a scalar function and the functions $\Gamma_n$ satisfy the following cell problems 
\begin{subequations}
\label{eq:Gamma_eqs}
\begin{align}
 \sigma_{ij} \sigma_{ik} \dfrac{\partial^2 \Gamma_n}{\partial y_j\partial y_k} = 0, \qquad &\bm{y} \in \omf(x_1), \\ 
 \sigma_{ij} \sigma_{ik} n_j^y \dfrac{\partial \Gamma_n}{\partial y_k} = \sigma_k n_k^y, \qquad &\bm{y} \in \pa \oms({x_1}), \\ 
 \Gamma_n \quad \text{ periodic on} \quad &\bm{y}\in\partial\omega_{=}\ \ \text{and}\ \ \partial\omega_{||}. 
\end{align}
Note that we enforce
\begin{equation}
 \langle\Gamma_n\rangle = 0,
\end{equation}
\end{subequations}
which uniquely defines $\Gamma_n$. Equations~\eqref{eq:Gamma_eqs} are obtained by substituting Equation \eqref{eq:Gamma} into Equation \eqref{eq:c_1storder}. Equations (\ref{eq:Gamma_eqs}) must then be solved numerically for $n\in\{1,2\}$ and each desired cell geometry (\textit{i.e.,} pairs of $a$ and $\lambda$). Note that Equations (\ref{eq:Gamma_eqs}) are independent of $\bm{\nabla}_x c^{(0)}$, justifying our scalings in Equation~\eqref{eq:Gamma}.

The goal of this analysis remains to determine a macroscale equation for the concentration. Since there are no macroscopic transport mechanisms present at this order, there is not enough information to determine a macroscale governing equation for the concentration. Hence, we must proceed to the next order in Equations (\ref{ad_diff_scaled}), which yield 
\begin{subequations}
\label{eq:c_2nsorder}
\begin{align}
\label{dc0dt}
\frac{\pa c^{(0)}}{\pa t} = \sigma_{ij}\dfrac{\partial \mathcal{A}_i}{\partial y_j} + \dfrac{\partial \mathcal{B}_i}{\partial x_i}, \qquad &\bm{y} \in \omf(x_1), \\
 \label{BC_adsorp}
 -\gamma c^{(0)} \left[\sigma_{kl}\sigma_{km} n^y_ln^y_m \right]^{1/2} = \sigma_{ij}n_j^y \mathcal{A}_i +
 N_i \mathcal{B}_i 
 \qquad &\bm{y} \in \pa \oms({x_1}), \\ 
 \quad{v}^{(2)}_i,\quad c^{(2)} \quad\text{periodic on} \quad &\bm{y}\in\partial\omega_{=}\ \ \text{and}\ \ \partial\omega_{||},
\end{align}
where
\begin{align}
\mathcal{A}_i \defeq \sigma_{ij} \dfrac{\partial c^{(2)}}{\partial y_j} + \dfrac{\partial c^{(1)}}{\partial x_i} - \Pen \ \bra{ v^{(0)}_i c^{(1)} + v^{(1)}_i c^{(0)} }, \\
\mathcal{B}_i \defeq \sigma_{ij} \dfrac{\partial c^{(1)}}{\partial y_j} + \dfrac{\partial c^{(0)}}{\partial x_i} - \Pen \ v^{(0)}_i c^{(0)}.
\end{align}
\end{subequations}
Integrating Equation (\ref{dc0dt}) over the transformed microscale fluid domain $\omega_f$ gives 
\begin{equation}
\label{dcdt_deriv}
|\omega_f| \frac{\partial c^{(0)}}{\partial t} = \int_{\omega_f} \! \sigma_{ij}\dfrac{\partial \mathcal{A}_i}{\partial y_j} \, \mathrm{d}S_y + \int_{\omega_f} \! \dfrac{\partial \mathcal{B}_i}{\partial x_i} \, \mathrm{d}S_y.
\end{equation}
Applying the divergence theorem to the first integral on the right-hand side of Equation (\ref{dcdt_deriv}) yields 
\begin{equation}\label{junk}
\int_{\omega_f} \! \sigma_{ij}\dfrac{\partial \mathcal{A}_i}{\partial y_j} \, \mathrm{d}S_y = \int_{\partial\omega_s} \! \sigma_{ij}n^y_j \mathcal{A}_i\, \mathrm{d}s_y+\int_{\partial\omega} \! \sigma_{ij}n^{\square}_j \mathcal{A}_i\, \mathrm{d}s_y,
\end{equation}
where $\mathrm{d}s_y$ signifies an element of a scalar line integral, and $\bm{n}^{\square} = n^{\square}_j \bm{e}_j$ is the outward-facing unit normal to the external square boundary $\partial\omega$. Since $\mathcal{A}_i$ is periodic on $\partial\omega$, the last term on the right-hand side of Equation~\eqref{junk} vanishes. Then, using Equation~(\ref{BC_adsorp}), we may re-write Equation~\eqref{junk} as
\begin{equation}\label{A_stuff}
\int_{\omega_f} \! \sigma_{ij}\dfrac{\partial \mathcal{A}_i}{\partial y_j} \, \mathrm{d}S_y =
-\int_{\partial\omega_s} \! N_i \mathcal{B}_i \, \mathrm{d}s_y -\int_{\partial\omega_s} \! \gamma c^{(0)}\left[\sigma_{kl}\sigma_{km} n^y_ln^y_m \right]^{1/2} \, \mathrm{d}s_y.
\end{equation}

To manipulate the final integral on the right-hand side of Equation (\ref{dcdt_deriv}), we apply the transport theorem (Eq.~\ref{transport_final}):
\begin{equation}
\label{B_stuff}
\int_{\omega_f} \! \dfrac{\partial \mathcal{B}_i}{\partial x_i} \, \mathrm{d}S_y = \dfrac{\partial}{\partial x_i} \int_{\omega_f} \! \mathcal{B}_i \, \mathrm{d}S_y + \int_{\partial \omega_s}\! N_i \mathcal{B}_i \,\mathrm{d}s_y.
\end{equation}
Thus, combining Equations (\ref{dcdt_deriv}), (\ref{A_stuff}) and (\ref{B_stuff}) we obtain
\begin{equation}
\label{phiCt}
|\omega_f| \frac{\partial c^{(0)}}{\partial t} = \dfrac{\partial}{\partial x_i} \int_{\omega_f} \! \left[\sigma_{ij} \dfrac{\partial c^{(1)}}{\partial y_j} + \dfrac{\partial c^{(0)}}{\partial x_i} - \Pen \ v^{(0)}_i c^{(0)}\right] \, \mathrm{d}S_y - \gamma c^{(0)} \int_{\partial\omega_s} \!\left[\sigma_{kl}\sigma_{km} n^y_ln^y_m \right]^{1/2} \, \mathrm{d}s_y. 
\end{equation}

Using the definitions of $c^{(1)}$ (Eq.~\ref{eq:Gamma}) and $\bm{V}^{(0)}$ (Eq.~\ref{V_exp}) and dividing through by $|\omega_f| = \phi$, we can re-write Equation~\eqref{phiCt} as 
\begin{subequations}
\label{final}
\begin{align}
\label{eq:c_homogenised}
\frac{\partial C^{(0)}}{\partial t}= \dfrac{1}{|\omega_f| }\dfrac{\partial}{\partial \hat{x}_i}\left[|\omega_f| D_{ij}(\phi,a) \dfrac{\partial C^{(0)}}{\partial \hat{x}_j} - \Pen \ |\omega_f| V^{(0)}_i C^{(0)} \right] - \gamma F(\phi,a) C^{(0)},
\end{align}
where we have expanded the intrinsic concentration $C$ in powers of $\epsilon$
\begin{equation}
\label{C_exp}
C(\bm{x},t) = C^{(0)}(\bm{x},t) + \eps C^{(1)}(\bm{x}, t) + \eps^2 C^{(2)}(\bm{x},t) +\cdots \quad \mbox{as} \quad \eps \to 0.
\end{equation}
and have noted that $C^{(0)}= c^{(0)}$. Note also that we have replaced $x_i$ with $\hat{x}_i$ in Equation~\eqref{eq:c_homogenised} since has been averaged over the microscale and is thus independent of $\bm{y}$. In Equation~\eqref{eq:c_homogenised}, the components of the effective diffusivity tensor $D_{ij}(\phi, a)$ are defined as 
\begin{align}
\label{eq:eff_D}
D_{ij}(\phi, a) \defeq \delta_{ij} - \frac{1}{|\omf|} \int_{\omf} \! \sigma_{ij} \dfrac{\partial \Gamma_j}{\partial y_j} \, \mathrm{d}S_y = \delta_{ij} - \frac{1}{\phi} \int_{\omf} \! \sigma_{ij} \dfrac{\partial \Gamma_j}{\partial y_j} \, \mathrm{d}S_y, 
\end{align}
where $\delta_{ij}$ is the Kronecker delta, and the effective adsorption strength $F(\phi, a)$ is defined as
\begin{align}
\label{source}
F(\phi,a) \defeq \dfrac{1}{|\omega_f|} \int_{\partial\omega_s} \!\left[\sigma_{kl}\sigma_{km} n^y_ln^y_m \right]^{1/2} \, \mathrm{d}s_y. 
\end{align}
\end{subequations}

Hence, our homogenized transport equation is given by Equations~\eqref{eq:c_homogenised}, \eqref{eq:eff_D} and (\ref{source}). In order to interpret the coefficients in this equation physically and evaluate them numerically, we now transform our coefficients into the naive microscale.

\subsubsection{Transforming into the naive microscale coordindate}

To interpret the rate $F(\phi,a)$ physically, it is helpful to map its definition Equation~\eqref{source} to the naive microscale coordinate $\bm{Y}$, defined in Equation~\eqref{Y_map}, in a similar way to \citet{richardson2011derivation}. Firstly, consider an arbitrary vector function $\bm{z}(\bm{x},\bm{y},t) = \bm{z}^\star(\bm{x},\bm{Y},t)$, such that $\bm{z} = z_i\bm{e}_i$ and $\bm{z}^\star = z_i^\star\bm{e}_i$ then Equation (\ref{volume_int_y2Y}), with $Z = \bm{\nabla}_y^a\cdot\bm{z} \equiv \bm{\nabla}_Y\cdot\bm{z}^\star = Z^\star$ gives 
\begin{equation}
\label{todiv}
 \frac{1}{|\omega(x_1)|}\int_{\omega(x_1)} \ \sigma_{ij}\frac{\partial z_i}{\partial y_j}, \mathrm{d} S_y \equiv \frac{1}{|\omega^{\star}(x_1)|}\int_{\omega^{\star}(x_1)} \ \frac{\partial z^\star_i}{\partial Y_i} \, \mathrm{d}S_Y, 
\end{equation} 
Thus taking the Divergence theorem of both sides of Equation (\ref{todiv}) leads to the relation 
\begin{align}
\label{eq: Jac calc general}
\dfrac{1}{|\omega|}\int_{\partial\omega_s} \!\sigma_{ij} z_j n^y_i \, \mathrm{d}s_y = \dfrac{1}{|\omega^{\star}|} \int_{\partial\omega_s^{\star}} \!{n^Y_i}^\star z^\star_i \, \mathrm{d}s_Y,
\end{align}
where ${n}_i^Y(\bm{y}) = {n_i^{Y}}^\star(\bm{Y})$. Note that as $\bm{\sigma}$ is diagonal $\sigma_{ij} z_j n^y_i = \sigma_{ij} n^y_jz_i$.  Additionally, Equations (\ref{eq: normals}) lead to the relation 
\begin{align}
\label{eq: normal equiv}
 \sigma_{ij} n^y_j = n^{Y}_{i} \left[\sigma_{kl} \sigma_{km} n_l^y n_m^y \right]^{1/2}.
\end{align}
Thus, setting $z_i=n_i^Y$ gives 
\begin{equation}
 \int_{\partial\omega_s} \left[\displaystyle \sigma_{kl}\sigma_{km} n^y_ln^y_m \right]^{1/2}\mathrm{d}s_y =\int_{\partial\omega_s} \sigma_{ij} n^y_jn_i^Y\mathrm{d}s_y = \frac{|\omega|}{|\omega^\star|}\int_{\partial\omega_s^\star}\ \mathrm{d}s_Y = \frac{|\partial\omega_s^\star|}{|\omega^\star|},
\end{equation}
since $\bm{n}^Y\cdot\bm{n}^Y = \bm{n}^{Y^\star}\cdot\bm{n}^{Y^\star}= 1$ and $|\omega|=1$. Hence,
\begin{equation}
\label{eq: F transform}
 F(\phi,a) = \frac{|\partial\omega_s^\star|}{|\omega_f||\omega^\star|}\equiv \frac{|\partial\omega_s^\star|}{|\omega_f^\star|},
\end{equation}
and we deduce that $F$ represents the obstacle perimeter within a cell, normalised by the fluid area within a cell.

Additionally, in order to evaluate $D_{ij}$ we find it convenient to map the system (Eq.~\ref{eq:Gamma_eqs}) to the naive microscale coordinate $\bm{Y}$, defined in Equation~\eqref{Y_map}. This mapping transforms the cell problems (Eq.~\ref{eq:Gamma_eqs}) to 
\begin{subequations}
\label{eq:Gamma_eqs scaled}
\begin{align}
\nabla^2_{Y} \Gamma_n^\star = 0, \qquad &\bm{Y} \in \omf^{\star}(x_1), \\ 
{\bm{n}^Y}^\star \cdot \nabla_{Y} \Gamma_n^\star = {n_k^Y}^\star, \qquad &\bm{Y} \in \pa \oms^{\star}({x_1}), \\ 
\Gamma_n^\star \quad \text{ periodic on} \quad &\bm{Y}\in\partial\omega_{=}^{\star}\ \ \text{and}\ \ \partial\omega_{||}^{\star},
\end{align}
with 
\begin{equation}
\langle\Gamma_n^\star\rangle=0. 
\end{equation}
\end{subequations}
\begin{subequations}
\label{macro_C}
The components of the effective diffusivity tensor (Eq.~\ref{eq:eff_D}) become 
\begin{align}
\label{eq:eff_D trans}
D_{ij}(\phi, a) = \delta_{ij} - \frac{1}{|\omf^{\star}|} \int_{\omf^{\star}} \! \dfrac{\partial \Gamma_j^\star}{\partial Y_i} \, \mathrm{d}S_Y = \delta_{ij} - \frac{1}{a \phi} \int_{\omf^{\star}} \! \dfrac{\partial \Gamma_j^\star}{\partial Y_i} \, \mathrm{d}S_Y.
\end{align}
We can also write Equation~\eqref{eq:eff_D trans} in tensor form $\bm{D}$ as 
\begin{align}
\bm{D}(\phi, a) = \II - \frac{1}{|\omf^{\star}|} \int_{\omf^{\star}} \! \grad{Y}\otimes{\GG}^\star \, \mathrm{d}S_Y.
\end{align} 
To evaluate $\bm{D}$, we solve the transformed cell problems (Eq.~\ref{eq:Gamma_eqs scaled}) numerically in COMSOL Multiphysics\textsuperscript{\textregistered}. 

Using the results from this subsection, we may re-write Equation~\eqref{eq:c_homogenised} in vector/tensor form as
\begin{align}
\label{eq:c_homogenised final}
\frac{\partial C^{(0)}}{\partial t}= \dfrac{1}{\phi} \hat{\nabla} \cdot \left(\phi \bm{D} \hat{\nabla} C^{(0)} - \Pen \ \phi \bm{V}^{(0)} C^{(0)}\right) - \gamma \dfrac{|\partial\omega_s^{\star}|}{|\omega_f^{\star}|} C^{(0)}. 
\end{align}
Again, since there is no $\bm{y}$-dependence in Equation (\ref{eq:c_homogenised final}), we have replaced $\bm{\nabla}_x$ with $\hat{\bm{\nabla}}$. 
\end{subequations}
Equation \eqref{eq:c_homogenised final} describes macroscopic transport by advection and diffusion in a porous medium with chemical sorption, where $\phi\bm{D}\cdot\hat{\bm{\nabla}}C^{(0)}$ is the diffusive flux per unit area of porous medium and $\bm{D}\cdot\hat{\bm{\nabla}}C^{(0)}$ is the diffusive flux per unit area of fluid. The form of the effective macroscale transport equation (Eq.~\ref{eq:c_homogenised final}) is similar to that obtained in \citet{dalwadi2015understanding}, where a simpler problem with a constant cell size is considered, resulting in a more straightforward upscaling procedure. Here, we have formally accounted for a slowly varying cell size and a slowly varying microscale geometry. The most significant difference between structure of the macroscale Equation~\eqref{eq:c_homogenised final} and the equivalent result obtained from the classic homogenisation of a strictly periodic problem arises from the slowly varying microscale geometry, which manifests through the explicit (nontrivial) porosity dependence of the diffusive term (\textit{cf.} \citet{bruna2015diffusion}). We find that the effect of the slowly varying cell size is less important to the structural form of the derived macroscale equations, which was not known at the outset.
 
Here, we have formally accounted for slowly varying cell size and slowly varying obstacle size. The resulting model has the same form as that derived in previous work for uniform cell size and slowly varying obstacle size (\textit{cf.} \citet{bruna2015diffusion}), in the sense that there are three macroscopic coefficients $\bm{K}$, $\bm{D}$, and $F$ that vary with the local microstructure via the values of phi, $R$, and here also $a$. Allowing for slowly varying cell size has not otherwise altered the mathematical structure of the macroscale problem at leading order, suggesting that different types of microscale heterogeneity can lead to a similar mathematical structure on the macroscale. The key difference between these heterogeneous results and the classical result for a uniform microstructure is that factors of porosity appear in front of the time derivative and within the divergence, the latter multiplying the macroscopic solute flux.

In \S\ref{figs}, we calculate the permeability and effective diffusivity for a porous medium with a simple, prescribed microstructure, we graphically present the resulting $\bm{D}(\phi, a)$, and we discuss its implications for this case. Note that although our model problem of a one-dimensional filter in \S\ref{sec:1Dfilter} features flow in the longitudinal direction, the macroscopic flow and transport equations (Eqs.~\ref{solvability}), (Eqs.~\ref{Incomp_macro}) and (Eqs.~\ref{eq:c_homogenised final}) and the results in \S\ref{general_K_D_stuff} are valid for any arbitrary flow direction.

\section{Illustrative example} \label{sec:effective_quantities}

\label{figs}
\begin{figure}
\centering
\includegraphics[width=\textwidth]{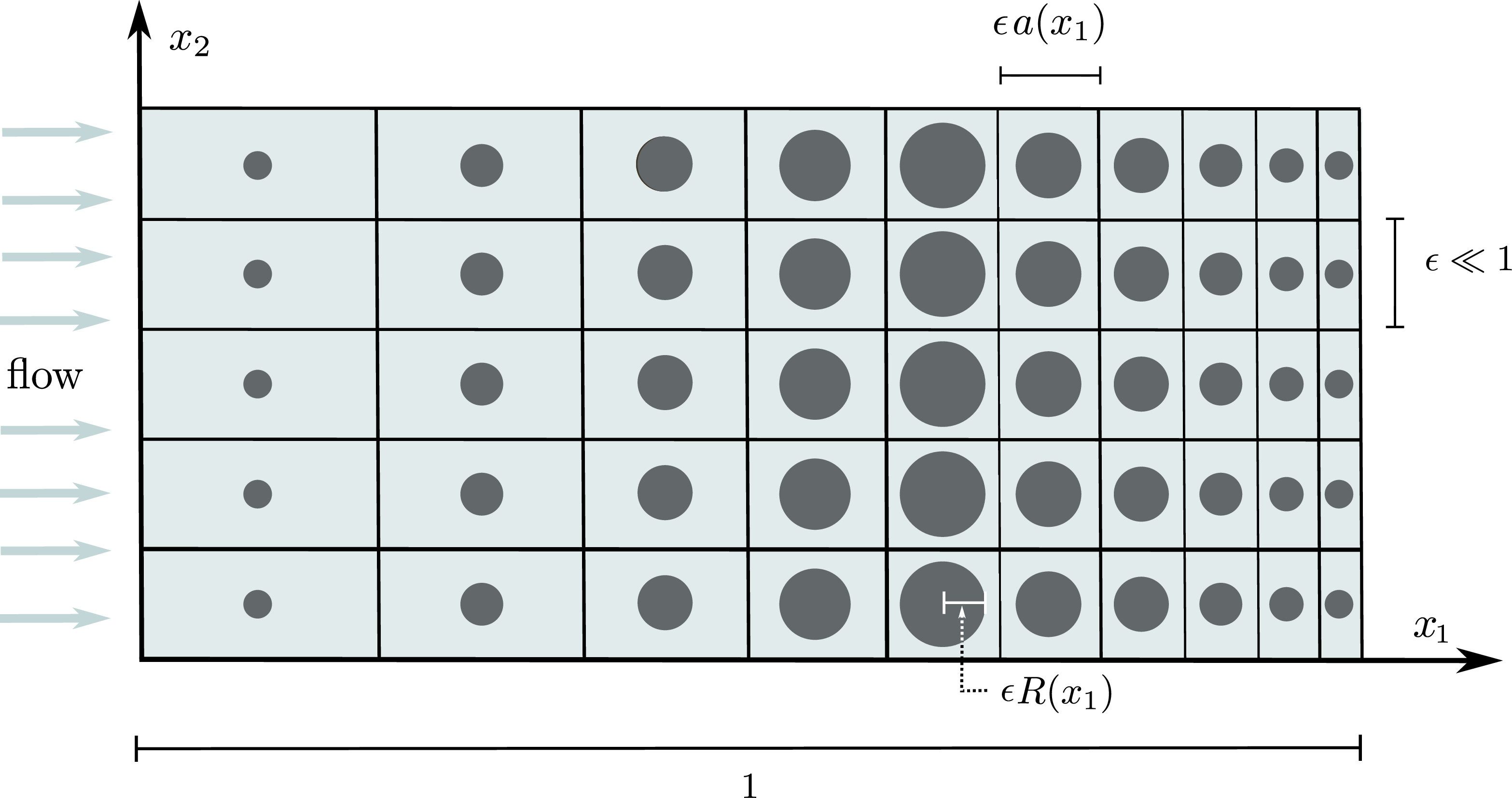} %Fig_Schem_circles_rect_lattice_dimensionless.eps
\caption{ \label{fig:schem_dim} We consider the flow of fluid carrying solute through a heterogeneous porous material in two dimensions for a specific illustrative example. Here, the porous medium is of unit length and is formed of an array of circular obstacles of dimensionless radius $R({x}_1)$, each located in the centre of a rectangular cell of unit transverse height and longitudinal width $a({x}_1)$.}
\end{figure}
\begin{figure}
\centering
\includegraphics[scale=1]{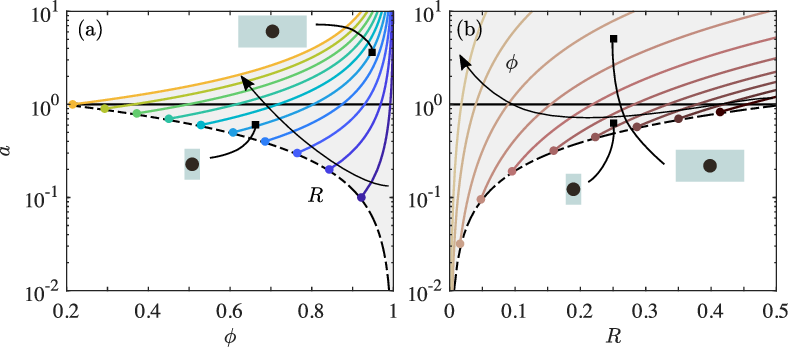}%as_vs_phis_LCA_1x2.eps
\caption{The porosity $\phi$ of a rectangular cell increases with aspect ratio $a$ and decreases with obstacle radius $R$ according to Equation~\eqref{phi_a_R}. (a)~$a$ versus $\phi$ for $R\in\{0.05,0.1,0.15,0.2,0.25,0.3,0.35,0.4,0.45,0.5\}$ (dark to light). The attainable region of the $\phi$-$a$ plane (shaded grey) is bounded below by $\phi_\text{min}$ given by Equation (\ref{phi}) with $a=2R$ (dot-dashed line). The cell geometry for two distinct points with $R= 0.25$ is shown pictorially. (b)~$a$ versus $R$ for $\phi\in\{0.35, 0.45, 0.55, 0.65, 0.75, 0.85, 0.925, 0.975, 0.995, 0.999\}$ (dark to light). The attainable region of the $R$--$a$ plane (shaded grey) is also bounded below by $a_\text{min}\defeq2R$ (dot-dashed line). Note that the smallest attainable $\phi$ for any $R$, $a$ combination is $\phi_\text{min}(R=1/2)=1-\pi/4$. \label{fig:AR_phi} }
\end{figure}

In this section, we examine a specific pore structure where the solid domain constitutes an array of solid circular obstacles centred on a rectangular lattice. Specifically, each cell contains a fixed, rigid circular obstacle of dimensionless radius $R({x}_1)$ at its centre. Since $R(x_1)$ uniquely controls the obstacle size over the length of the medium, we take the scale factor $\lambda(x_1) = R(x_1)$. To prevent the obstacles from overlapping, we require that $2R\leq{}\mathrm{min}(a,1)$. 
This construction leads to a porous medium whose properties vary in the longitudinal direction but not in the transverse direction (see Figure~\ref{fig:schem_dim}). For this geometry the porosity $\phi$ is
\begin{equation}
\label{phi_a_R}
\phi(x_1) = \frac{|\omega_f(x_1)|}{|\omega(x_1)|} = \frac{|\omega_f^{\star}(x_1)|}{|\omega^{\star}(x_1)|} \equiv
1-\frac{\pi R(x_1)^2}{a(x_1)}, 
\end{equation}
since $|\omega^{\star}| = a$ and $|\omega_f^{\star}| = a - \pi R^2$. Further, in this case we may explicitly evaluate the effective adsorption rate $F(\phi,a)$ in Equation~\eqref{source} using the formulation from Equation~\eqref{eq: F transform}, giving
\begin{equation}
\label{source_rect}
F(\phi,a) = \dfrac{|\partial \omega^{\star}_s|}{|\omega^{\star}_f|} = \frac{2 \pi R}{a \phi } = \frac{2\left(1-\phi\right)}{R \phi}. 
\end{equation}
Note that with this geometry and in the limit $a=1$, Equations~\eqref{final} become the same system as Equation~(3.22) in \citet{dalwadi2015understanding} in two dimensions (\textit{i.e.,} $d=2$), but written in terms of the intrinsic average 
rather than the volumetric average.

\subsection{Macroscale flow and transport properties}
\label{general_K_D_stuff}

For this specific geometry we explore the impact of microstructure on macroscopic flow and transport by analysing the permeability and effective net diffusivity tensors, $\bm{K}$ and $\phi\bm{D}$, respectively. To determine $\bm{K}$ we solve Equations \eqref{stokes_homog_O_ep_scaled_naive} in COMSOL Multiphysics\textsuperscript{\textregistered} using the `Laminar Flow (spf)' interface (`Fluid Flow' $\to$ `Single Phase Flow' $\to$ `Laminar flow (spf)'). The domain is discretised using the `Physics-controlled mesh' with the element size set to `Extremely fine'. Similarly, to evaluate $\bm{D}$, we solve Equations \eqref{eq:Gamma_eqs scaled}, in COMSOL Multiphysics\textsuperscript{\textregistered} using the `Laplace Equation (lpeq)' interface (`Classical PDEs' $\to$ `Mathematics branch' $\to$ `Laplace Equation (lpeq)'). For the flow problem, the domain is discretised using the `Physics-controlled mesh' with the element size set to `Extremely fine'. 

\begin{figure}
\centering
\includegraphics[width=\textwidth]{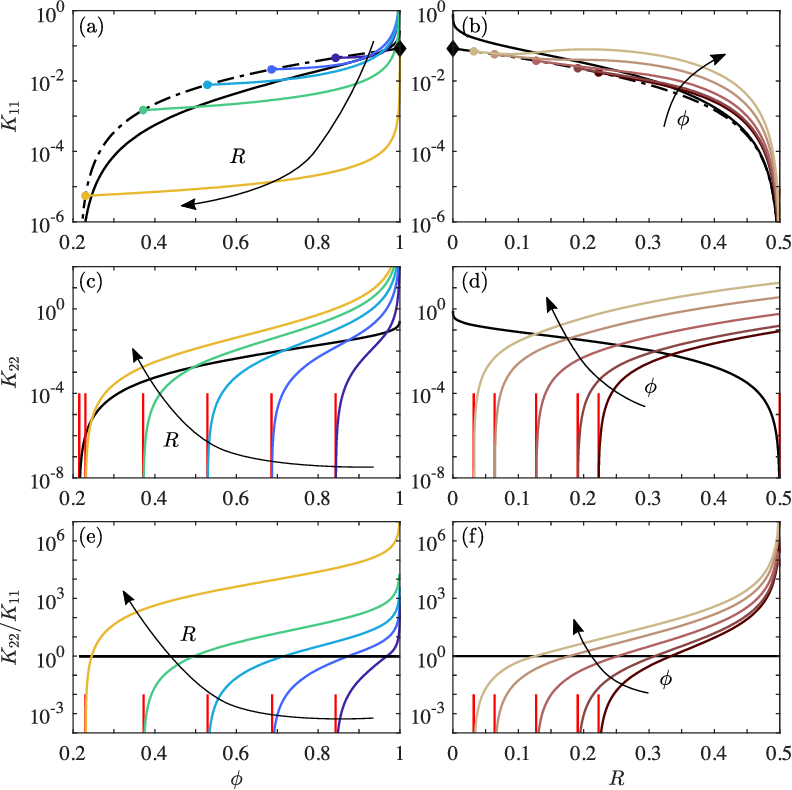}%K_vs_phi_V2_LCA_3x2.eps
\caption{The longitudinal permeability $K_{11}$ (top row), transverse permeability $K_{22}$ (middle row) and the permeability--anisotropy ratio $K_{22}/K_{11}$ (bottom row) depend strongly on microstructure. Left column:~$K_{11}$, $K_{22}$ and $K_{22}/K_{11}$ against $\phi$ for fixed values of $R\in\{0.1, 0.2, 0.3, 0.4, 0.49\}$, with $a$ varying according to Equation~\eqref{phi_a_R}. Right column:~the same quantities against $R$ for fixed values of $\phi\in\{0.65, 0.7, 0.8, 0.9, 0.95\}$, with $a$ varying according to Equation~\eqref{phi_a_R}. For a given value of $R$, the minimum porosity $\phi_\mathrm{min}(R)$ is given by Equation~\eqref{phi_a_R} with $a=2R$. Note that $K_{11}$ is non-zero at $\phi_\mathrm{min}(R)$ for all values of $\phi$ (dot-dashed curve, top row), whereas $K_{22}$ vanishes at $\phi_\mathrm{min}$ ((c) and (e) red vertical asymptotes). There exists a smallest possible $R$ for any given $\phi$ ((d) and (f) red vertical asymptotes). In all cases, $K_{11}$ and $K_{22}$ are as defined in Equation \eqref{K_def} and calculated using COMSOL Multiphysics\textsuperscript{\textregistered}. The permeability is isotropic when $a\equiv1$ (solid black curves; \citet{dalwadi2015understanding}). The limit $R\to0$ and $a \to0$ corresponds to a set of parallel but disconnected channels with unit transverse width, for which $\phi\to1$, $K_{22}\to0$, and $K_{11}\to1/12$ (black diamonds in top row). \label{fig:perm} }
\end{figure}

\begin{figure}
\centering
\includegraphics[width=\textwidth]{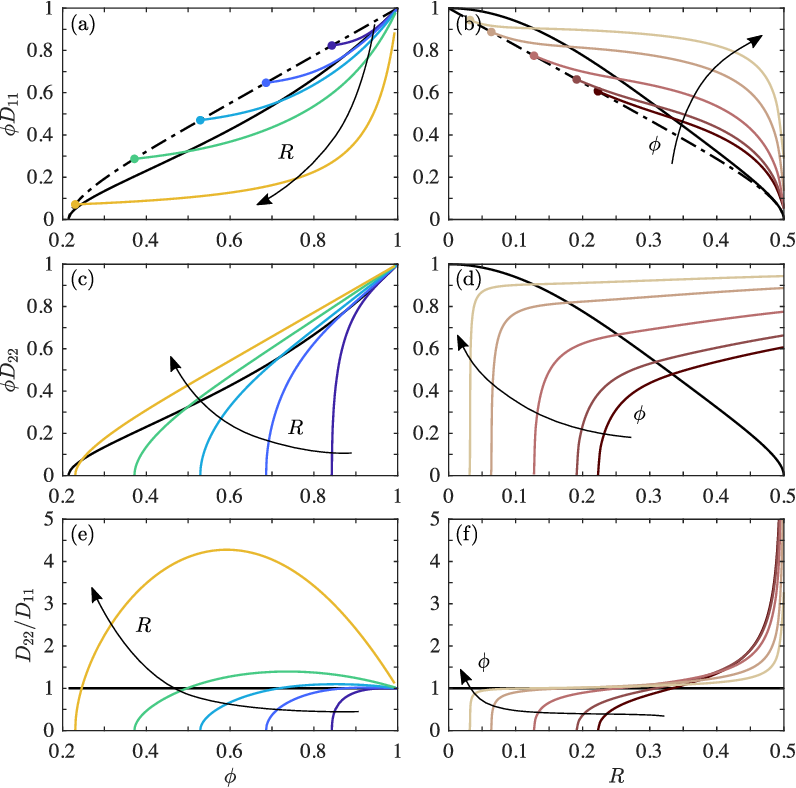}%D_vs_phi_LCA_3x2_new.eps
\caption{The net longitudinal diffusivity $\phi D_{11}$ (top row), net transverse diffusivity $\phi D_{22}$ (middle row) and the diffusivity--anisotropy ratio $D_{22}/D_{11}$ (bottom row) depend strongly on microstructure. 
Left column:~$\phi D_{11}$, $\phi D_{22}$ and $D_{22}/D_{11}$ against $\phi$ for fixed values of $R\in\{0.1, 0.2, 0.3, 0.4, 0.49\}$, with $a$ varying according to Equation~\eqref{phi_a_R}. Right column:~the same quantities against $R$ for fixed values of $\phi\in\{0.65, 0.7, 0.8, 0.9, 0.95\}$, with $a$ varying according to Equation~\eqref{phi_a_R}. For a given value of $R$, the minimum porosity $\phi_\text{min}(R)$ is given by Equation~\eqref{phi_a_R} with $a=2R$. Note that $\phi D_{11}$ is non-zero at $\phi_\mathrm{min}(R)$ for all values of $\phi$ (dot-dashed curve, top row), whereas $\phi D_{22}$ vanishes at $\phi_\text{min}$. In all cases, $D_{11}$ and $D_{22}$ are as defined in Equation~\eqref{eq:eff_D trans} and calculated using COMSOL Multiphysics\textsuperscript{\textregistered}. Both $D_{11}$ and $D_{22}$ tend to 1 as $\phi\to1$, which corresponds to the limit of free-space diffusion. The net effective diffusivity, $\phi\bm{D}$, is isotropic when $a=1$ (solid black curves), in agreement with the results presented in \citet{dalwadi2015understanding}. \label{fig:diff} }
\end{figure}

The tensors $\bm{K}$ and $\bm{D}$ depend on microstructure via $a$, $\phi$ and $R$, any two of which are independent and the third prescribed by Equation (\ref{phi_a_R}) (Figure~\ref{fig:AR_phi}). We therefore have one additional degree of microstructural freedom relative to \citet{dalwadi2015understanding} and this allows us to explore the anisotropy in the system. 
We explore the effect of the $a$, $R$ and $\phi$ parameter space on $\bm{K}$ and $\phi\bm{D}$ in Figures \ref{fig:perm} and \ref{fig:diff}, respectively. The effective diffusivity $\bm{D}$ is shown for reference in Figure~\ref{fig:diffAppendix} (top and middle row; Appendix~\ref{s:diffAppendix}).

\begin{figure}
\centering
\includegraphics[width=\textwidth]{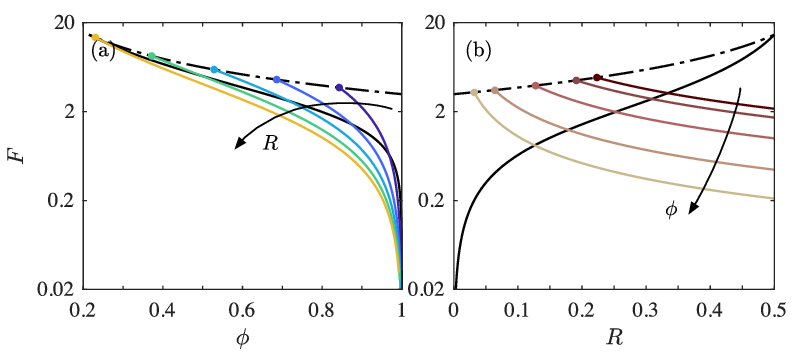}%fig_F_arrows.eps
\caption{The effective adsorption rate $F$ depends strongly on microstructure. Left column: $F$ against $\phi$ for fixed values of $R\in\{0.1, 0.2, 0.3, 0.4, 0.49\}$, with $a$ varying according to Equation (\ref{phi_a_R}). Right column: $F$ against $R$ for fixed values of $\phi\in\{0.65, 0.7, 0.8, 0.9, 0.95\}$, with $a$ varying according to Equation (\ref{phi_a_R}). In all cases, $F$ is as defined in Equation~\eqref{source_rect}. For a given value of $R$, the minimum porosity $\phi_\text{min}(R)$ is given by Equation (\ref{phi_a_R}) with $a=2R$. The results of \citet{dalwadi2015understanding} are again reproduced when $a=1$ (solid black). \label{fig:F} }
\end{figure}

We have validated our analysis for this geometry in a number of ways. Firstly, we have compared our results with those in \citet{dalwadi2015understanding} for the special case of $a\equiv 1$, confirming both the final homogenised equations (Eqs.~\ref{solvability}, \ref{eq:eff_D trans}, \ref{eq:c_homogenised final}, and \ref{source_rect}) and the detailed numerical results (black lines; Figures \ref{fig:perm}--\ref{fig:F}). Secondly, we have confirmed our results in the Hele-Shaw limit of parallel, disconnected channels where the longitudinal permeability is $1/12$ (black diamond; Figure~\ref{fig:perm}) and the transverse permeability vanishes. Finally, we have confirmed that both the transverse permeability and the transverse effective diffusivity vanish when the transverse connectivity vanishes ($a\to{}2R$ or $2R\to{}a$; red lines in Figures~\ref{fig:perm} and \ref{fig:diff}), and that the permeability diverges and the effective diffusivity tends to unity as the obstacles vanish ($\phi\to 1$).

Increasing $\phi$ at fixed $R$ is achieved by increasing $a$ (Figure \ref{fig:AR_phi}~(a)), such that the obstacles move further apart in the longitudinal direction only; as a result, $K_{11}$, $K_{22}$, $\phi D_{11}$ and $\phi D_{22}$ all increase (Figure \ref{fig:perm}~(a) and (c) and Figure \ref{fig:diff}~(a) and (c)). As $\phi\to1$ ($a\to\infty$), both $K_{11}$ and $K_{22}$ diverge as the resistance to flow vanishes (Figure \ref{fig:perm}~(a) and (c)), and both $\phi D_{11}$ and $\phi D_{22}$ tend to 1 as molecular diffusion becomes unobstructed (Figure \ref{fig:diff} (a) and (c)). As $\phi\to\phi_\mathrm{min}(R)$ ($a\to2R$) at fixed $R$, the obstacles move closer together in the longitudinal direction and the pore space becomes disconnected in the transverse direction, so that $K_{22}$ and $\phi D_{22}$ vanish; $K_{11}$ and $\phi D_{11}$ are minimised but do not vanish. Further taking $R\to0$, the longitudinal problem reduces to a set of disconnected parallel channels of unit transverse width, for which $K_{11}=1/12$ (Figure \ref{fig:perm} top row, black diamond).

Increasing $R$ at fixed $\phi$ is similarly achieved by increasing $a$ (Figure \ref{fig:AR_phi}~(b)), in which case the transverse channels between obstacles grow wider while the longitudinal channels between obstacles grow narrower. As a result, $K_{11}$ and $\phi D_{11}$ decrease while $K_{22}$ and $\phi D_{22}$ increase. As $R \to 1/2$ at fixed $\phi$, the longitudinal channels close and $K_{11}$ and $\phi D_{11}$ vanish, but the transverse channels become wider and $K_{22}$ and $\phi D_{22}$ are maximised. The longitudinal permeability, $K_{11}$, is weakly non-monotonic in $R$ for larger values of $\phi$(Figure~\ref{fig:perm}~(b)), which means that the longitudinal permeability of a high porosity porous material can be maximised for a given $\phi$ by appropriately varying $R$ and $a$.

When $a\equiv1$, equivalent to the case considered in \citet{dalwadi2015understanding}, $\bm{K}$ and $\phi\bm{D}$ become isotropic. Increasing $\phi$ corresponds to decreasing $R$, in which case both the longitudinal and transverse spacing between obstacles decreases (Figure \ref{fig:AR_phi}) which decreases $K_{11}=K_{22}$ and $\phi D_{11}=\phi D_{22}$ (Figures \ref{fig:perm} and \ref{fig:diff}, solid black lines). For $a\not=1$, our microstructure is inherently anistropic ($K_{11}\neq K_{22}$, $\phi D_{11}\neq \phi D_{22}$). For $a<1$, the longitudinal channels are wider than the transverse channels, such that $K_{22}/K_{11}<1$ and $D_{22}/D_{11}<1$ and both ratios vanish as $a\to2R$ ($\phi\to\phi_\text{min}$ where $\phi_\text{min}$ is given by Equation (\ref{phi_a_R}); Figure \ref{fig:perm}~(e) and (f) and Figure \ref{fig:diff}~(e) and (f), respectively). 
For $a>1$, the longitudinal channels are narrower than the transverse channels, such that $K_{22}/K_{11}> 1$ and $D_{22}/D_{11}>1$. The permeability--anisotropy ratio, $K_{22}/K_{11}$, increases monotonically with both $\phi$ and $R$, and diverges as $\phi\to1$ at fixed $R$ ($K_{22}$ diverges faster than $K_{11}$ because the obstacles never get further apart in the transverse direction) and as $R\to1/2$ at fixed $\phi$ ($K_{11}$ vanishes; Figure~\ref{fig:perm}~(e)~and~(f)). 
The diffusivity--anisotropy ratio, $D_{22}/D_{11}$, increases monotonically with $R$ for all $\phi\in(\phi_\text{min},1)$, diverging as $R\to1/2$ (Figure \ref{fig:diff}~(f)). For fixed $R$, this ratio increases monotonically with $\phi$ for $a\leq1$, is equal to unity for $a=1$ (isotropic geometry), and must approach unity as $\phi\to1$ ($a\to\infty$; unobstructed molecular diffusion; Figure \ref{fig:diff}(e)). These bounds require that $D_{22}/D_{11}$ has an intermediate maximum in $\phi$ (or in $a$) at fixed $R$, the amplitude of which diverges as $R\to1/2$. Specifically, the non-monotonicity in the ratio $D_{22}/D_{11}$ occurs due to the relative rates of increase of $\phi D_{11}$ and $\phi D_{22}$. For $\phi=\phi_\text{min}$ ($a = 2R$) there is no transverse connectivity thus separating the obstacles slightly (a small increase in $a$) leads to a sharp increase in $\phi D_{22}$ but only a slight increase in $\phi D_{11}$ since the longitudinal connectivity is unchanged and most longitudinal mixing occurs in the longitudinal channels. Conversely, as $\phi\to1$ ($a\to\infty$) the longitudinal spacing between obstacles diverges which means that $\phi D_{11}$ is very sensitive to changes in $a$ as longitudinal mixing occurs predominantly between longitudinally adjacent obstacles (in the transverse channels), thus $\phi D_{11}$ approaches unity rapidly. However, significant transverse connectivity is preserved for large $a$ so increasing $a$ further has minimal effect on $\phi D_{22}$ since most transverse mixing occurs in the transverse channels in this limit. Note that the longitudinal diffusivity $D_{11}$ is non-monotonic in $\phi$ for each $R$ as $a$ varies (Appendix \ref{s:diffAppendix}, Figure \ref{fig:diffAppendix}).

The partially absorbing boundary condition on the microscale, whose strength is measured by the parameter ${\gamma}$ in Equation~\eqref{Robin}, leads to an effective sink term in the macroscale transport problem, whose strength is measured by $\gamma F$, where $F$ is given in Equation~\eqref{source_rect}. $F$ is the ratio of the perimeter of an obstacle to the fluid area within a cell, which are $2\pi{}R$ and $a \phi$, respectively, for a rectangular array of circular obstacles. We consider the impact of microstructure on the removal of solute in more detail in \S\ref{sec:1Dfilter}. Note that $F$ decreases as $\phi$ increases at fixed $R$, as should be expected, but also as $R$ increases at fixed $\phi$; the latter occurs because an increase in obstacle size requires a correspondingly larger increase in cell size to keep $\phi$ constant. We consider the impact of microstructure on the removal of solute in more detail in \S\ref{sec:1Dfilter}.

\subsection{Simple one-dimensional filter}\label{sec:1Dfilter}

We now use the homogenised model to understand the effect of microstructure and P\'{e}clet number on filter efficiency in the context of a simple one-dimensional steady-state filtration problem. We identify the performance of the filter with the rate at which it removes solute, and thus use the leading-order outlet concentration $C_{\textrm{out}}^{(0)}$ as a measure of filtration efficiency. Specifically, we consider Equations~(\ref{solvability}) and (\ref{final}) at steady state, with imposed flux and concentration at the inlet, 
\begin{subequations}
\label{BCs_macro_filter_dim}
\begin{align}
\label{BCs_macro_filter}
\phi{\bm{V}^{(0)}} = \bm{e}_1 \quad &\text{at} \quad \hat{x}_1 = 0,\\
\label{Cin}
C^{(0)} = 1 \quad &\text{at} \quad \hat{x}_1 = 0,
\end{align}
and passive outflow at the outlet,
\begin{equation}
\label{Cout}
\hspace{-2mm} \frac{\partial C^{(0)}}{\partial \hat{x}_1} = 0 \quad \text{at} \quad \hat{x}_1 = 1.
\end{equation}
\end{subequations}
Since these boundary conditions (Eq.~\ref{BCs_macro_filter_dim}a) are compatible with unidirectional flow, we take $\bm{V}^{(0)}(\hat{\bm{x}})=V_1^{(0)}(\hat{x}_1)\bm{e}_1$ and $C^{(0)}(\hat{\bm{x}},t) = C^{(0)}(\hat{{x}}_1)$. Thus, Equation~\eqref{Incomp_macro} leads to 
\begin{equation}\label{Incomp_1D}
 \frac{\mathrm{d}}{\mathrm{d}\hat{x}_1} \bra{\phi V_1^{(0)}} = 0,
\end{equation}
which, on application of the inlet condition (Eqs.~\ref{BCs_macro_filter}), gives the macroscale flux $\phi{V}_1^{(0)} \equiv 1$ for all $\hat{x}_1$. The associated pressure drop across the entire filter, $\Delta P^{(0)}$, is obtained by integrating Equation~\eqref{solvability} and using the fact that $\phi{V}_1^{(0)} \equiv 1$, which gives
\begin{align}
\label{effective permeability}
 \Delta P^{(0)} = \int_0^1 \frac{1}{K_{11}(x)}\;\mathrm{d}x.
\end{align}
Note that the right-hand side of Equation~\eqref{effective permeability} is a measure of the total flow resistance of the entire filter; the inverse of this quantity can be thought of as an effective permeability for the entire filter.

Hence, the homogenised governing equation for the steady concentration distribution $C^{(0)}(\hat{x}_1)$ (Eq.~\ref{eq:c_homogenised final}) becomes
\begin{equation}\label{eq:BVP1}
 \dfrac{1}{\phi}\frac{\mathrm{d}}{\mathrm{d} \hat{x}_1} \sqbra{ \phi D_{11}(\phi, R) \frac{\mathrm{d} C^{(0)}}{\mathrm{d} \hat{x}_1} - \Pen \ C^{(0)} } = \gamma F(\phi, R) C^{(0)} \qquad \mbox{for} \quad \hat{x}_1 \in (0, 1), 
\end{equation}
where $F(\phi, R)$ is defined in Equation (\ref{eq: F transform}). We solve Equation (\ref{eq:BVP1}) subject to Equations~\eqref{Cin} and \eqref{Cout} numerically using a finite-difference scheme. We discretise the interval [0, 1] using a uniform mesh of size $\Delta x = 1/N$ and we approximate derivatives using a second-order accurate central-difference formula. The results presented below were obtained using $N = 500$. Below, we consider filters with varying porosity and filters with uniform porosity, but with heterogeneous microstructure in both cases.

\subsubsection{Porosity gradients} 

\begin{figure}
 \centering
 \includegraphics[scale=1]{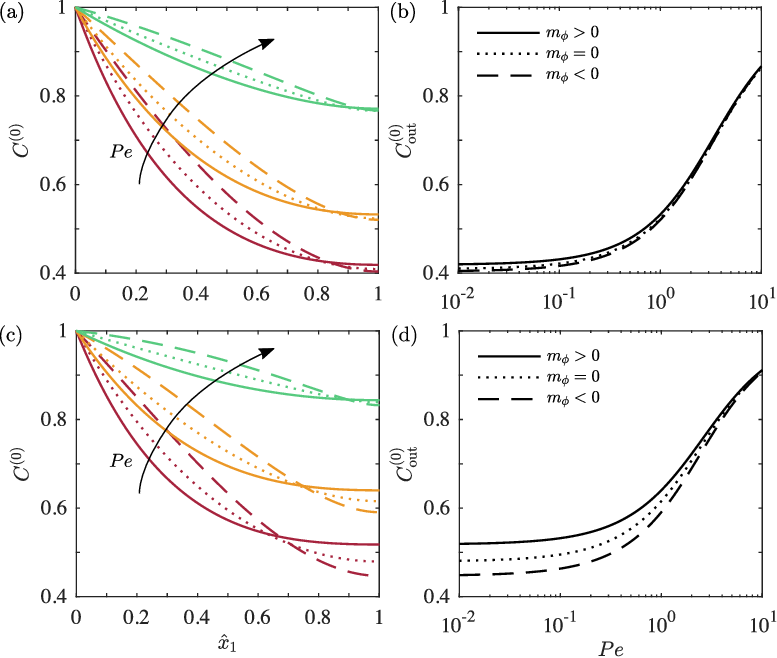} %C_phi_grad.eps
 \caption{Steady-state concentration field $C^{(0)}(\hat{x}_1)$ for $\Pen \in\{ 0, 1, 5\}$ (left column) and outlet concentration $C_{\rm out}^{(0)}\defeq C^{(0)}(1)$ as a function of $\Pen$ (right column). Top row:~$a\equiv1$ and $\phi(\hat{x}_1) = 0.8 + m_\phi (\hat{x}_1 - 0.5)$. Bottom row:~$R\equiv0.4$ and $\phi(\hat{x}_1) = 0.8 + m_\phi (\hat{x}_1 - 0.5)$. In both cases, $m_\phi = 0.3$ (solid), $m_{\phi} = 0$ (dotted), and $m_\phi = -0.3$ (dashed). \label{fig7} }
\end{figure}

We first consider filters with varying porosity. Recall that a given porosity $\phi$ may be achieved in two different ways: by fixing $a(\hat{x}_1)$ and varying $R(\hat{x}_1)$, as considered in \citet{dalwadi2015understanding} for $a\equiv1$; or by fixing $R(\hat{x}_1)$ and varying $a(\hat{x}_1)$. We consider these options in Figure \ref{fig7}, for the same three porosity fields in both cases: linearly increasing with $\hat{x}_1$, uniform in $\hat{x}_1$, or linearly decreasing with $\hat{x}_1$. Specifically, we take $\phi(\hat{x}_1) = \phi_0 + m_\phi(\hat{x}_1-0.5)$, where $\phi_0$ is the average porosity (also the mid-point porosity) and $m_\phi$ is the porosity gradient. We take $\phi_0=0.8$ and $m_\phi=0.3$ (increasing in $\hat{x}_1$), $m_\phi=0$ (uniform in $\hat{x}_1$), or $m_\phi=-0.3$ (decreasing in $\hat{x}_1$). Note that, $|m_\phi|$ defines the filter microstructure and $\mathrm{sgn}(m_\phi)$ is simply the orientation of the filter.

When varying $\phi$ by varying $R$ at fixed $a\equiv1$, as considered by \citet{dalwadi2015understanding}, the sign of the porosity gradient has a modest impact on the concentration distribution within the filter: $m_\phi>0$ leads to a steeper gradient in $C^{(0)}$ near the inlet and a shallower gradient in $C^{(0)}$ near the outlet, whereas $m_\phi<0$ leads to a more uniform gradient in $C^{(0)}$ throughout the filter (Figure \ref{fig7}(a)). However, the outlet concentration $C_\text{out}^{(0)}\defeq C^{(0)}(1)$ is remarkably insensitive to $m_\phi$.The outlet concentration is slightly lower for $m_\phi<0$, and this slight difference decreases as $\Pen$ increases (Figure \ref{fig7}(b)). As $\Pen$ increases, advection becomes stronger causing more solute to be swept through the filter; as a result, $C^{(0)}(\hat{x}_1) $ increases with $\Pen$ for all $\hat{x}_1$, and $C^{(0)}_\mathrm{out}$ more than doubles as $\Pen$ increases from 0 to 10. The case when $a\equiv1$ is considered in more detail in \citet{dalwadi2015understanding}. Varying $\phi$ by varying $a$ at fixed $R\equiv0.4$ leads to qualitatively similar results, but $C^{(0)}(x)$ and $C^{(0)}_\text{out}$ are more sensitive to $m_\phi$ (Figure \ref{fig7}, bottom row). For all $\Pen$, attaining a desired porosity gradient via varying $R$ leads to a more efficient filter than varying $a$, in the sense that $C_\text{out}^{(0)}$ is lower for the same $\phi(x_1)$. 

\subsubsection{Microstructural gradients with uniform porosity}
\label{const_phi}

\begin{figure}
 \centering
 \includegraphics[scale=1]{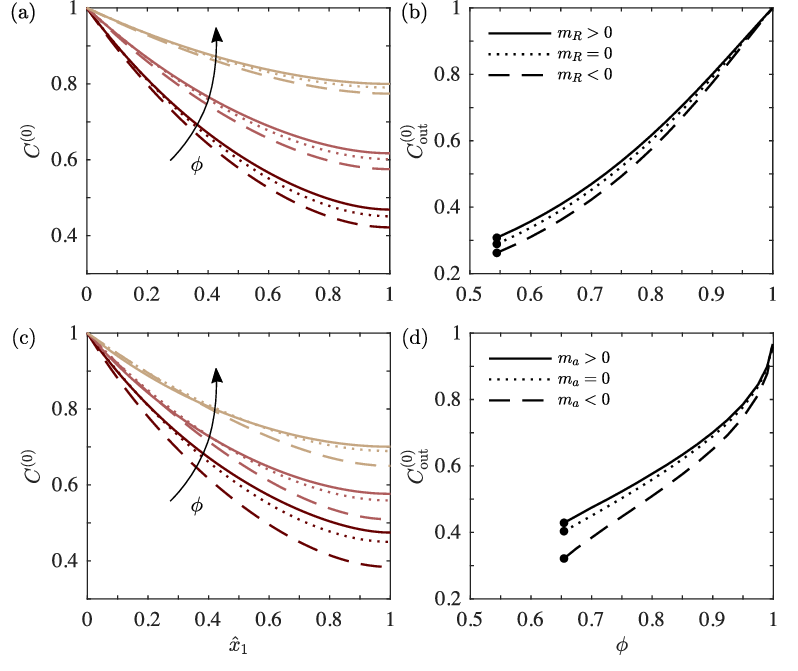} %C_const_phi.eps
 \caption{ Steady-state concentration field $C^{(0)}(\hat{x}_1)$ for $\phi \in \{0.7, 0.8, 0.9\}$ (left column), and outlet concentration $C_{\rm out}^{(0)}\defeq C^{(0)}(1)$ as a function of $\phi$ (right column). Top row: $\Pen = 1$ and $R(\hat{x}_1) = 0.36 + m_R(\hat{x}_1-0.5)$ for $m_R = 0.14$ (solid), $m_{R} = 0$ (dotted), $m_R = -0.14$ (dashed). The aspect ratio $a$ varies following Equation~\eqref{phi_a_R}. For $R_0=0.36$ and $m_R=\pm0.14$ the minimum attainable porosity is $1-0.145\pi\approx0.54$ ((b); black dots). Bottom row: $\Pen = 1$ and $a(\hat{x}_1) = 1.34 + m_a(\hat{x}_1-0.5)$ for $m_a = 1.8$ (solid), $m_{a} = 0$ (dotted), and $m_a = -1.8$ (dashed). The radius $R$ varies following Equation~\eqref{phi_a_R}. For $a_0=1.34$ and $m_a=\pm1.8$ the minimum attainable porosity is $1-0.11\pi\approx0.66$ ((d); black dots). \label{fig8} }
\end{figure}

We now consider filters with uniform porosity but gradients in microstructure. We therefore fix $\phi$ and simultaneously vary $R$ and $a$ with $\hat{x}_1$, recalling that $a$ is related to $R$ via Equation~\eqref{phi_a_R}. We consider two types of variation: an imposed gradient in $R$ with $a$ varying to maintain constant $\phi$ (via Equation~\eqref{phi_a_R}) or an imposed gradient in $a$ with $R$ varying to maintain constant $\phi$ (via Equation~\eqref{phi_a_R}). We consider these options in Figure \ref{fig8}. 

We first consider $R(\hat{x}_1) = R_0 + m_R(\hat{x}_1-0.5)$, where $R_0$ is the average obstacle radius (also the mid-point) radius and $m_R$ is the gradient (Figure~\ref{fig8}, top row). We take $R_0 = 0.36$ and $m_R=0.14$ (increasing in $\hat{x}_1$), $m_R=0$ (uniform in $\hat{x}_1$) or $m_R=-0.14$ (decreasing in $\hat{x}_1$). 

When varying $R$ linearly, the sign of $m_R$ has a modest impact on the concentration distribution: for any uniform porosity, $m_R>0$ leads to a shallower gradient in $C^{(0)}$ and a higher concentration at every point within the filter, including the outlet. 
Thus, for any uniform porosity, $m_R<0$ is always a more efficient filter than $m_R>0$. 

We next consider $a(\hat{x}_1) = a_0 + m_a(\hat{x}_1-0.5)$, where $a_0$ is the average cell width (also the cell width at the mid-point) and $m_a$ is the gradient of $a$ over $\hat{x}_1$ (Figure \ref{fig8} bottom row). 
We take $a_0 = 1.34$ and $m_a=1.8$ (increasing in $\hat{x}_1$), $m_a=0$ (constant in $\hat{x}_1$), or $m_a=-1.8$ (decreasing in $\hat{x}_1$). From Equation~\eqref{phi_a_R} for fixed $\phi$, it can be seen that $a \propto R^2$, thus, a decrease in $a$ must be mirrored by a decrease in $R$ to maintain a uniform $\phi$. Thus, we expect the same qualitative behaviour for a linear gradient in $R$ (Figure~\ref{fig8}, top row) as for a linear gradient in $a$ (Figure~\ref{fig8}, bottom row). We find that $m_a<0$ leads to a more efficient filter for all $\phi$. 

Note that, for any $\phi\in(1-0.11\pi,1)$ --- that is, the range of porosities attainable for imposed linear gradients in both $R$ and $a$ (see Figure \ref{fig8} caption) --- prescribing $a$ and taking $m_a<0$ predicts most efficient filter considered here (comparing Figure \ref{fig8}~(b) and Figure \ref{fig8}~(d)). Similarly, for large $\phi\gtrsim0.725$ prescribing $a$ and taking $m_a>0$ predicts the least efficient filter, whereas, for small $\phi\lesssim 0.725$ prescribing $R$ and taking $m_R>0$ predicts the least efficient filter. 

\section{Conclusions} \label{sec:conclusion}

We have systematically derived a macroscopic model for flow, transport and sorption during steady flow in a two-dimensional heterogeneous and anisotropic porous medium using generalisations of standard homogenisation theory for slow variations in the size of periodic cells \citep{chapman2011unified,richardson2011derivation} and locally periodic microstructures \citep{bruna2015diffusion, dalwadi2015understanding}. We derived a model valid for a heterogeneous porous medium comprising cells of varying size each containing multiple arbitrarily shaped obstacles. The heterogeneity originates from slowly varying obstacle size and/or obstacle spacing along the length of the porous medium, the latter also induces strong anisotropy within the problem. For the flow problem, we obtain Darcy's law with an anisotropic permeability tensor, and for the solute concentration problem we obtain an advection–diffusion–reaction equation with an anisotropic effective diffusivity tensor. The permeability, effective diffusivity and the removal term are functions of the porosity, obstacle spacing and a scale factor controlling the variation in obstacle size across the medium; any two of these are free choices which prescribe the third. In \S\ref{sec:effective_quantities} we consider a simple geometry comprising a circular obstacle centred in a rectangular cell. We determine the corresponding permeability and effective diffusivity numerically and show how this depends on the obstacle radius and aspect ratio of the rectangle. This work illustrates and quantifies how the permeability and diffusivity of a porous medium not only depend on the porosity of the medium, but also depend strongly on the microstructure of the medium.

The homogenisation procedure we used allows for slowly varying changes to the cell surrounding each circle that comprises the filter. This means that the total area of each individual cell may differ between cells. This is a new aspect to homogenisation and we have carefully derived a transport theorem to account for how these microstructural changes affect the macroscale transport. Using this transport theorem we have shown that macroscale incompressibility is preserved (the divergence of the Darcy flux vanishes) and that this is independent of the individual cell size. The two degrees of microstructural freedom (varying obstacle size and spacing) enable us to consider a wide range of heterogeneities on the microscale, for example, to maintain a uniform porosity while systematically varying the microstructure. These macroscale equations are computationally inexpensive to solve, allowing for optimisation of parameters through large sweeps, which would not be possible with direct numerical simulations.

We have focused on a regime in which diffusion balances advection and removal at the macroscale and dominates advection and removal at the microscale (\textit{i.e.}, $Pe=O(1)$, $\epsilon{}Pe\ll1$). Sub-limits involving weaker advection and/or removal may be taken directly in the final result without repeating the interim analysis. For scenarios with stronger advection (\textit{i.e.}, $\epsilon{}Pe=O(1)$), as might be the case in many industrial filtration scenarios, hydrodynamic dispersion becomes important and new terms that are proportional to the product of velocity and concentration gradient will arise in the homogenised equations. Our analysis here lays the foundation for future work to incorporate dispersive effects.

The example geometry considered in \S\ref{sec:effective_quantities} is two-dimensional; a direct physical analogue would be a quasi-two-dimensional filter comprising solid circular pillars that are centered on a rectangular grid and sufficiently tall that boundary effects at the top and bottom walls can be neglected. This is a simple but appropriate model for non-woven fibrous filters, which form a major part of the filtration industry (\textit{e.g.,} those in air purifiers and vacuum cleaners)~\citep{spychala2015bacteria, printsypar2019influence}, magnetic separation filters composed of wire wool~\citep{mariani2010high}, and microfluidic devices containing tall micropillars~\citep{benitez2012microfluidic, wang2013ciliated}. The strong anisotropy in the problem could be useful for filter design; it is achieved while maintaining the circular shape of the obstacles and the principal directions of the permeability and diffusivity tensors are fixed as the longitudinal and transverse directions. Furthering our understanding of the impacts of microstructural heterogeneity and anisotropy in general, is of use to many other areas of research including hydrology and biology (\textit{e.g.,}\citealp[]{wang2020effect, o2015multiscale}).

We considered a simple model problem for a one-dimensional filter with chemical adsorption at steady state. Measuring efficiency as the amount of solute removed by the filter per unit time, we found that negative porosity gradients lead to a more efficient filter than positive porosity gradients or filters of uniform porosity. Further, for a fixed porosity, decreasing obstacle size or decreasing obstacle spacing lead to more efficient filters than their respective constant or increasing counterparts. For a given porosity decreasing the obstacle spacing linearly leads to a more efficient filter than linearly decreasing the obstacle radius. 

While we have defined efficiency to mean instantaneous performance, there are further considerations to a filter's efficiency. Factors such as manufacturing costs, filter lifetime and fluid flux output may also need to be considered. For example, if the coating on the solid obstacles was very expensive then we may wish to minimise the amount of surface area of the solid obstacles while maximising performance. Further, we assumed that the solid surface never saturates with solute. However, in practice, the number of active sites where the solute can attach to the solid will decrease as solute adsorbs, which may reduce the efficiency. In this case, this effect may be mitigated by ensuring that there are active sites throughout the full length of the filter so that the chance a solute particle comes into contact with an active site is maximised. The simple one-dimensional filter model we considered only predicts initial or instantaneous filter efficiency and will therefore not predict the total amount of contaminant filtered out over the life span of a filter if properties were to change with time. However, the equations derived in this paper can readily be generalised to describe such a case. All of these additional considerations to filter design lead to multiple optimisation problems, requiring large parameter sweeps, for which a computationally inexpensive model, such as this, is vital. 

In our analysis we have assumed that the solute particles are negligibly small; for particles that are not negligible in size relative to the smallest distances between adjacent obstacles (choke points), we would also need to consider the effects of choking of the filter due to particle build-up. Avoidance of such filter blockages requires sufficiently wide longitudinal connectivity. Hence, a filter comprising obstacles whose radii increase with depth is desirable, since such a gradient allows for more build-up of solute on the solid obstacles near the inlet without choking the filter. This scenario was considered in \citet{dalwadi2016multiscale}. However, in our case, we also have the possibility of varying the spacing between obstacles. This additional degree of freedom allows us to respect a positive gradient in the obstacle radii to mitigate the risks of blockages, while also having either a negative gradient in the porosity or obstacle spacing to enable more efficient filters.

We have validated our results against limiting cases and previous homogenisation results (\textit{cf.} \S\ref{general_K_D_stuff}); DNS for flow and transport in a broader range of relevant geometries would provide further validation and may lead to additional insight, and should be the subject of future work.

While it has been shown that the effective diffusivity for a porous medium with obstacles on a uniform square grid was qualitatively similar to a porous medium with obstacles on a uniform hexagonal grid for all porosities \citep{bruna2015diffusion}, we expect that the addition of anisotropy to the hexagonal problem, obtained by varying the longitudinal obstacle spacing, will cause the permeabilities and effective diffusivites to diverge from those determined here. For example, in certain limits, the hexagonal problem reduces to a series of longitudinal channels while in other limits, the hexagonal problem reduces to a series of transverse channels. Consequently in the latter limit, for the hexagonal structure, the longitudinal permeability and diffusivity must vanish, while for the rectangular structure the longitudinal permeability and diffusivity remain non-zero for the all parameter combinations. In general, the hexagonal structure of obstacles will mean that the longitudinal permeability will be more sensitive to longitudinal obstacle spacing than it is for obstacles in a rectangular structure. This is because with a hexagonal grid, altering the longitudinal spacing alters both the longitudinal and transverse distances between neighbouring obstacles, while for a rectangular grid, altering the longitudinal spacing does not alter the transverse distance between obstacles. This illustrates that, when anisotropy is introduced into a problem, the microstructure becomes more significant than for isotropic problems.

It would be straightforward to generalise our approach to a three-dimensional porous medium comprising spherical obstacles centred on a cuboid grid that is homogeneous in two directions, but again allowing for arbitrary variation of both obstacle radius and obstacle spacing in the longitudinal direction. We would expect that the results would be qualitatively similar to the two-dimensional problem considered here, however connectivity does not vanish when obstacles touch. This would then mean that we have non-zero permeability and diffusivity in all directions throughout the entire parameter space. 

A final point to note is that the spacing between obstacles may change when a filter is subject to an effective stress. By coupling the model presented here to a law that relates the spacing of the obstacles to the strain of the porous medium, we can derive homogenised equations for a filter undergoing longitudinal deformation. Modelling the filter as a series of circles on a varying hexagonal grid will better describe granular materials and this is the focus of future work. 

In summary, the results presented in this manuscript form a comprehensive framework for describing the transport and adsorption properties through heterogeneous porous media. The model can be used to answer questions on the filtration performance of such porous media as well as being well-equipped for the generalisation to more complicated scenarios. 

\section*{Acknowledgements}

M.P.D. would like to acknowledge helpful discussions with Professor S.~J.~Chapman.

\section*{Funding}

This work was supported by the Royal Society (L.C.A., grant reference number \verb+ICA\R1\180098+),  (I.M.G., University Research Fellowship with grant reference number \verb+URF\R\191008+); the European Research Council (ERC) under the European Union's Horizon 2020 Programme (L.C.A., C.W.M., and S.P., grant number 805469); and IIT Gandhinagar (S.P., Research Initiation Grant (RIG)).

\section*{Declaration of interests}

The authors report no conflict of interest.

\section*{Data availability statement} The data that support the findings of this study are openly available in GitHub at https://github.com/satyajitpramanik/homogenization-jfm2021.

\section*{Author ORCID}

L. C. Auton, https://orcid.org/0000-0003-2871-9191; S. Pramanik, https://orcid.org/0000-0001-8487-3551; M. P. Dalwadi https://orcid.org/0000-0001-5017-2116; C. W. MacMinn https://orcid.org/0000-0002-8280-0743; I. M. Griffiths https://orcid.org/0000-0001-6882-7977. 

\appendix

\section{Transport theorem} 
\label{transport_thm}
 
\subsection{Generalised transport theorem}
\label{transport_thm_gen}
Firstly, we present a generalised form of the transport theorem which allows us to interchange $\bm{\nabla}_x$ with integration over a cell of arbitrary geometry. Consider the region $\alpha$ bounded by the surface $\partial{\alpha}$, and suppose that this region moves and/or deforms with time $t$. Denote the position of points on $\partial{\alpha}(t)$ by ${\bm{y}}^b(t)$. The Reynolds Transport Theorem states that
\begin{equation}\label{eq:RTT}
 \frac{\ \mathrm{d}}{\ \mathrm{d}t}\int_{\alpha(t)}\,\bm{\zeta}\,\ \mathrm{d}V =\int_{\alpha(t)}\,\frac{\partial{\bm{\zeta}}}{\partial{t}}\,\ \mathrm{d}V +\int_{\partial{\alpha}(t)}\,\left(\frac{\partial{\bm{y}^b}}{\partial{t}}\cdot\bm{n}\right)\bm{\zeta}\,\ \mathrm{d}S,
\end{equation}
for an arbitrary vector field $\bm{\zeta}(\hat{\bm{x}},t)$, where $\mathrm{d}V$ signifies a volume integral, $\mathrm{d}S$ signifies a surface integral, $\bm{n}$ is the outward normal to $\partial{\alpha}$ and the time derivative $\partial{\bm{y}^b}/\partial{t}$ can be identified as the local velocity of $\partial{\alpha}(t)$.

In Equation~\eqref{eq:RTT}, $t$ plays the role of an arbitrary scalar parameter. In other words, Equation~\eqref{eq:RTT} remains valid if we suppose that the region moves and/or deforms according to some other scalar parameter $\xi$, in which case we have that 
\begin{equation}\label{eq:RTT2}
\frac{\mathrm{d}}{\mathrm{d}\xi}\int_{\beta(\xi)}\!\bm{z}\, \mathrm{d}V =\int_{\beta(\xi)}\!\frac{\partial{\bm{z}}}{\partial{\xi}}\, \mathrm{d}V +\int_{\partial{\beta}(\xi)}\!\left(\frac{\partial{\bm{y}^b}}{\partial{\xi}}\cdot\bm{n}\right)\bm{z}\, \mathrm{d}S, 
\end{equation}
for an arbitrary vector field $\bm{z}(\hat{\bm{x}}, \xi)$, and where the domain $\beta$ is a function of $\xi$. Note that the derivative $\partial{\bm{y}^b}/\partial{\xi}$ can no longer be identified as a velocity in the traditional sense.

Now, consider several independent parameters as a vector $\boldsymbol{\xi}=\xi_i{\bm{e}}_i$, where we use the summation convention and $\bm{e}_i$ is the unit normal in the $i^{\textrm{th}}$ direction. The corresponding divergence with respect to this vector is then
\begin{equation}\label{eq:div1}
 \boldsymbol{\nabla}_{\xi} \cdot\int_{\beta(\boldsymbol{\xi})}\!\bm{z}\, \mathrm{d}V=\left({\boldsymbol{e}}_i\frac{\partial}{\partial{\xi_i}}\right)\cdot\int_{\beta(\boldsymbol{\xi})}\!(z_j{\bm{e}}_j)\, \mathrm{d}V =\frac{\partial}{\partial{\xi_i}}\int_{\beta(\boldsymbol{\xi})}\!z_i\, \mathrm{d}V.
\end{equation}
Equation~\eqref{eq:RTT2} provides the following expression for the right-hand side of Equation~\eqref{eq:div1}
\begin{equation}\label{eq:div2}
 \frac{\partial}{\partial{\xi_i}}\int_{\beta(\boldsymbol{\xi})}\!z_i\, \mathrm{d}V =\int_{\beta(\boldsymbol{\xi})}\!\frac{\partial{z_i}}{\partial{\xi_i}}\, \mathrm{d}V +\int_{\partial{\beta}(\boldsymbol{\xi})}\!\left(\frac{\partial{\bm{y}^b}}{\partial{\xi_i}}\cdot\bm{n}\right)z_i\, \mathrm{d}A.
\end{equation}
Returning to vector notation, we can rewrite this result as
\begin{subequations}
\label{gen_transport_a}
\begin{equation}
 {\boldsymbol{\nabla}}_{\xi} \cdot\int_{\beta(\boldsymbol{\xi})}\!\bm{z}\, \mathrm{d}V=\int_{\beta(\boldsymbol{\xi})}\!{\boldsymbol{\nabla}}_\xi\cdot\bm{z}\, \mathrm{d}V +\int_{\partial{\beta}(\boldsymbol{\xi})}\!\bm{n}\cdot\bm{G}\cdot\bm{z}\, \mathrm{d}A,
\end{equation}
where
\begin{equation}
\label{gen_transport_b}
 \bm{G} =\frac{\partial{y^b_i}}{\partial{\xi_j}}\,{\bm{e}}_i{\bm{e}}_j=\left(\frac{\partial{\bm{y}^b}}{\partial{\boldsymbol{\xi}}}\right)^\intercal=({\boldsymbol{\nabla}}_\xi\otimes\bm{y}^b)^\intercal,
\end{equation}
\end{subequations}
is the Jacobian of the dependence of $\partial{\beta}$ on $\boldsymbol{\xi}$. Thus, Equation (\ref{gen_transport_a}) defines the generalised transport theorem. 

\subsection{Relationship to the macroscale perturbation to the normal}

Applying the generalised Reynolds transport theorem (Eq.~\ref{gen_transport_a}) to a vector field $\bm{z}(\bm{x},\bm{y})$, over the periodic cell $\omega_f$, yields the following expression
\begin{equation}
\label{gen_transport_1}
{\boldsymbol{\nabla}}_x \cdot\int_{\omega_f(x_1)}\!\bm{z}\, \mathrm{d}S_y=\int_{\omega_f(x_1)}\,\boldsymbol{\nabla}_x\cdot\bm{z}\, \mathrm{d}S_y +\int_{\partial{\omega_s(x_1)}}\!\bm{n}^y\cdot\bm{G}\cdot\bm{z}\, \mathrm{d}s_y
 +\int_{\partial{\omega(x_1)}}\!\bm{n}^\square\cdot\bm{G}\cdot\bm{z}\, \mathrm{d}s_y,
\end{equation}
where $\bm{n}^\square$ is defined in relation to Equation (\ref{junk}). 

From periodicity, the final term on the right-hand side of Equation~\eqref{gen_transport_1} vanishes, \emph{i.e.}
\begin{equation}
\int_{\partial{\omega(x_1)}}\!\bm{n}^\square\cdot\bm{G}\cdot\bm{z}\, \mathrm{d}s_y = 0.
\end{equation}
Using the definition of $\bm{n}^y$ from Equation~\eqref{eq: n geo} and the definition of $\bm{G}$ from Equation~\eqref{gen_transport_b}, we obtain
\begin{equation}
\label{eq: n G orig} 
\bm{n}^y\cdot\bm{G} = \frac{\displaystyle\frac{\partial f_s}{\partial y_i} \frac{\partial y_i^b}{\partial x_j}}{\displaystyle|\bm{\nabla}_y f_s|}\bm{e}_j,
\end{equation}
using the summation convention and evaluated on $\bm{y} = \bm{y}^b(\bm{x})$, which is defined implicitly through
\begin{align}
\label{eq: yb def}
f_s(\bm{x},\bm{y}^b(\bm{x})) = 0.
\end{align}

Since $\bm{x} = x_j \bs{e}_j$ is a parameter in each unit cell, differentiating Equation~\eqref{eq: yb def} with respect to $x_j$ yields the relationship
\begin{equation}\label{eq: diff fs wrt x}
\frac{\partial f_s}{\partial x_j}+\frac{\partial f_s}{\partial y_i}\frac{\partial y_i^b}{\partial x_j} = 0,
\end{equation}
evaluated on $\bm{y} = \bm{y}^b(\bm{x})$. Substituting the relationship (Eq.~\ref{eq: diff fs wrt x}) into Equation~\eqref{eq: n G orig} implies that 
\begin{equation}
 \bm{n}^y\cdot\bm{G} = -\frac{\bm{\nabla}_x f_s}{|\bm{\nabla}_y f_s|} \equiv - \bm{N},
\end{equation}
using the definition of $\bm{N}$ in Equation (\ref{normal_N}). Thus, the transport theorem (Eq.~\ref{gen_transport_1}) becomes
\begin{equation}\label{transport_final}
 \int_{\omega_f} \! \bm{\nabla}_x\cdot\bm{z}\, \mathrm{d}S_y = \bm{\nabla}_x\cdot\int_{\omega_f} \! \bm{z}\, \mathrm{d}S_y + \int_{\partial\omega_s} \! \bm{N}\cdot\bm{z}\, \mathrm{d}s_y.
\end{equation} 

\section{Non-monotonicity of \textbf{\emph{D}$_{\textbf{11}}$}}
\label{s:diffAppendix}

\begin{figure}
\centering
\includegraphics[scale=1]{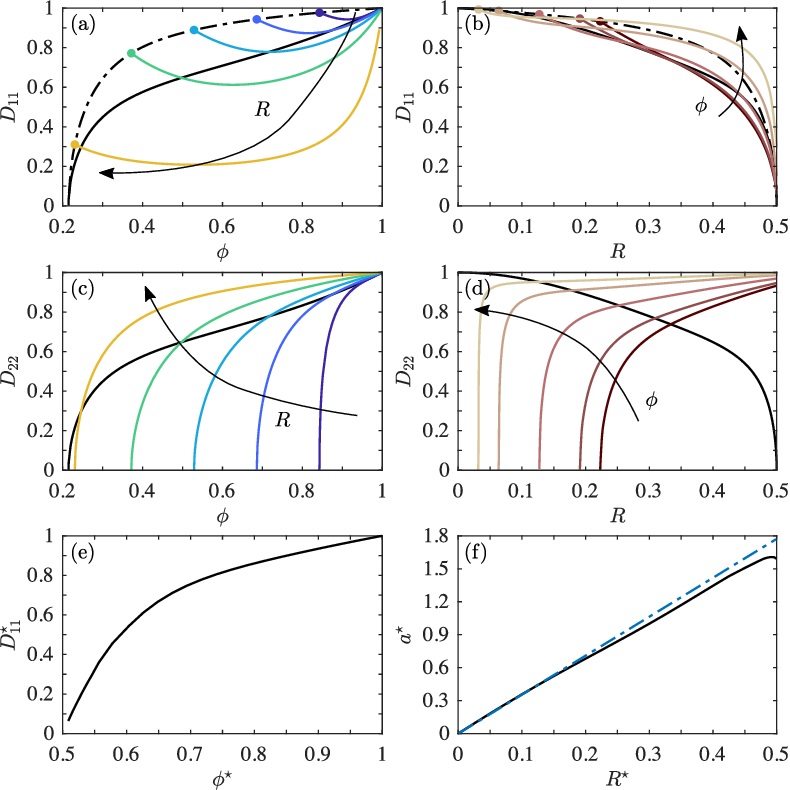} %D_vs_phi_LCA_3x2_appendix.eps
\caption{(a) and (c) $D_{11}$ and $D_{22}$, respectively, against $\phi$ for fixed values of $R\in\{0.1, 0.2, 0.3, 0.4, 0.49\}$, with $a$ varying according to Equation~\eqref{phi_a_R}. (b) and (d) $D_{11}$ and $D_{22}$, respectively, against $R$ for fixed values of $\phi\in\{0.65, 0.7, 0.8, 0.9, 0.95\}$, with $a$ varying according to Equation (\ref{phi_a_R}). For a given value of $R$, the minimum porosity $\phi_\text{min}(R)$ is given by Equation (\ref{phi_a_R}) with $a=2R$. Note that $D_{11}$ is non-zero at $\phi_\mathrm{min}(R)$ for all values of $\phi$ (dot-dashed curve, top row), whereas $D_{22}$ vanishes at the corresponding $\phi_\text{min}$.  Both $D_{11}$ and $D_{22}$ are as defined in Equation \eqref{eq:eff_D trans} and calculated using COMSOL Multiphysics\textsuperscript{\textregistered}. Both $D_{11}$ and $D_{22}$ tend to 1 as $\phi\to1$, which corresponds to the limit of free-space diffusion. The effective diffusivity is isotropic when $a=1$ (solid black curves, top 2 rows), in agreement with the results presented in \citet{dalwadi2015understanding}. (e)~The minimum longitudinal diffusivity $D_{11}=D_{11}^{\star}$ for each value of $R$ in Figure~\ref{fig:diff}(a) against the corresponding porosity $\phi=\phi^{\star}$ and (f) the corresponding value of $a=a^\star$ against $R^\star$. The latter is well predicted by the line $a^\star = 2 R^\star \sqrt{\pi}$ (blue dot-dashed line) for small $R^\star$, becoming non-monotonic near $R^\star=0.5$. \label{fig:diffAppendix} }
\end{figure}

In this Appendix, we show the diffusivity tensor $\bm{D}$ (Figure \ref{fig:diffAppendix}~(a)--(d)) and examine the non-monotonicity of the longitudinal diffusivity $D_{11}$ for fixed $R$ as $\phi$ varies. We consider the minimum longitudinal diffusivity $D_{11}^\star$ and the unique value of $\phi^\star$ to which it corresponds (Figure \ref{fig:diffAppendix}~(e)). Each value $\phi^\star$ corresponds to a particular pair $a^\star$ and $R^\star$ (Figure \ref{fig:diffAppendix}~(f)). Note that $a^\star$ is approximately related to $R^\star$ via $a^\star=2R^\star\sqrt{\pi}$ (Figure \ref{fig:diffAppendix}~(f), blue dot-dashed line).  This linear relationship is a good fit for small $R^\star$, but slightly overestimates the true value of $a^\star$ for larger values of $R^\star$.

\bibliographystyle{jfm}

\bibliography{Homog_refs} 

\begin{thebibliography}{42}
\expandafter\ifx\csname natexlab\endcsname\relax\def\natexlab#1{#1}\fi
\def\au#1{#1} \def\ed#1{#1} \def\yr#1{#1}\def\at#1{#1}\def\jt#1{\textit{#1}}
  \def\bt#1{#1}\def\bvol#1{\textbf{#1}} \def\vol#1{#1} \def\pg#1{#1}
  \def\publ#1{#1}\def\arxiv#1{#1}\def\org#1{#1}\def\st#1{\textit{#1}}

\bibitem[Auriault(1991)]{auriault1991heterogeneous}
{\sc \au{Auriault, J.-L.}} \yr{1991}  \at{Heterogeneous medium. {I}s an
  equivalent macroscopic description possible?}  \jt{International Journal of
  Engineering Science}  \bvol{29}~(7),  \pg{785--795}.

\bibitem[Beckwith {\em et~al.\/}(2003)Beckwith, Baird \&
  Heathwaite]{beckwith2003anisotropy}
{\sc \au{Beckwith, C.~W}, \au{Baird, A.~J.} \& \au{Heathwaite, A.~L.}}
  \yr{2003}  \at{Anisotropy and depth-related heterogeneity of hydraulic
  conductivity in a bog peat. {II}: modelling the effects on groundwater flow}.
   \jt{Hydrological processes}  \bvol{17}~(1),  \pg{103--113}.

\bibitem[Ben{\'\i}tez {\em et~al.\/}(2012)Ben{\'\i}tez, Topolancik, Tian,
  Wallin, Latulippe, Szeto, Murphy, Cipriany, Levy, Soloway \&
  Craighead]{benitez2012microfluidic}
{\sc \au{Ben{\'\i}tez, J.~J.}, \au{Topolancik, J.}, \au{Tian, H.~C.},
  \au{Wallin, C.~B}, \au{Latulippe, D.~R.}, \au{Szeto, K.}, \au{Murphy, P.~J.},
  \au{Cipriany, B.~R.}, \au{Levy, S.~L.}, \au{Soloway, P.~D.} \& \au{Craighead,
  H.~G.}} \yr{2012}  \at{Microfluidic extraction, stretching and analysis of
  human chromosomal {DNA} from single cells}.  \jt{Lab on a Chip}  \bvol{12},
  \pg{4848--4854}.

\bibitem[Bensoussan {\em et~al.\/}(2011)Bensoussan, Lions \&
  Papanicolaou]{bensoussan2011asymptotic}
{\sc \au{Bensoussan, A.}, \au{Lions, J.-L.} \& \au{Papanicolaou, G}} \yr{2011}
  {\em Asymptotic analysis for periodic structures\/}, ,  \vol{vol. 374}.
  \publ{American Mathematical Soc.}

\bibitem[Bruna \& Chapman(2015)]{bruna2015diffusion}
{\sc \au{Bruna, M.} \& \au{Chapman, S.~J.}} \yr{2015}  \at{Diffusion in
  spatially varying porous media}.  \jt{SIAM Journal on Applied Mathematics}
  \bvol{75}~(4),  \pg{1648--1674}.

\bibitem[Brusseau(1994)]{brusseau1994transport}
{\sc \au{Brusseau, M.~L.}} \yr{1994}  \at{Transport of reactive contaminants in
  heterogeneous porous media}.  \jt{Reviews of Geophysics}  \bvol{32}~(3),
  \pg{285--313}.

\bibitem[Chapman \& McBurnie(2011)]{chapman2011unified}
{\sc \au{Chapman, S.~J.} \& \au{McBurnie, S.~E.}} \yr{2011}  \at{A unified
  multiple-scales approach to one-dimensional composite materials and
  multiphase flow}.  \jt{SIAM Journal on Applied Mathematics}  \bvol{71}~(1),
  \pg{200--217}.

\bibitem[Chapman {\em et~al.\/}(2008)Chapman, Shipley \&
  Jawad]{chapman2008multiscale}
{\sc \au{Chapman, S.~J.}, \au{Shipley, R.~J.} \& \au{Jawad, R.}} \yr{2008}
  \at{Multiscale modeling of fluid transport in tumors}.  \jt{Bulletin of
  Mathematical Biology}  \bvol{70},  \pg{2334 -- 2357}.

\bibitem[Clavaud {\em et~al.\/}(2008)Clavaud, Maineult, Zamora, Rasolofosaon \&
  Schlitter]{clavaud2008permeability}
{\sc \au{Clavaud, J.-B.}, \au{Maineult, A.}, \au{Zamora, M.}, \au{Rasolofosaon,
  P.} \& \au{Schlitter, C.}} \yr{2008}  \at{Permeability anisotropy and its
  relations with porous medium structure}.  \jt{Journal of Geophysical
  Research: Solid Earth}  \bvol{113}~(B1).

\bibitem[Dalwadi {\em et~al.\/}(2016)Dalwadi, Bruna \&
  Griffiths]{dalwadi2016multiscale}
{\sc \au{Dalwadi, M.~P.}, \au{Bruna, M.} \& \au{Griffiths, I.~M.}} \yr{2016}
  \at{A multiscale method to calculate filter blockage}.  \jt{Journal of Fluid
  Mechanics}  \bvol{809},  \pg{264--289}.

\bibitem[Dalwadi {\em et~al.\/}(2015)Dalwadi, Griffiths \&
  Bruna]{dalwadi2015understanding}
{\sc \au{Dalwadi, M.~P.}, \au{Griffiths, I.~M.} \& \au{Bruna, M.}} \yr{2015}
  \at{Understanding how porosity gradients can make a better filter using
  homogenization theory}.  \jt{Proceedings of the Royal Society A}  \bvol{471},
   \pg{20150464}.

\bibitem[Daly \& Roose(2015)]{daly2015homogenization}
{\sc \au{Daly, K.~R.} \& \au{Roose, T.}} \yr{2015}  \at{Homogenization of two
  fluid flow in porous media}.  \jt{Proceedings of the Royal Society A}
  \bvol{471},  \pg{20140564}.

\bibitem[Davit {\em et~al.\/}(2013{\natexlab{{\em a\/}}})Davit, Bell, Byrne,
  Chapman, Kimpton, Lang, Leonard, Oliver, Pearson, Shipley, L., P., D. \&
  M.]{davit2013homogenization}
{\sc \au{Davit, Y.}, \au{Bell, C.~G.}, \au{Byrne, H.~M.}, \au{Chapman, L.
  A.~C.}, \au{Kimpton, L.~S.}, \au{Lang, G.~E.}, \au{Leonard, K. H.~L.},
  \au{Oliver, J.~M.}, \au{Pearson, N.~C.}, \au{Shipley, R.~J.}, \au{L.,
  Waters~S.}, \au{P., Whiteley~J.}, \au{D., Wood~B.} \& \au{M., Quintard}}
  \yr{2013{\natexlab{{\em a\/}}}}  \at{Homogenization via formal multiscale
  asymptotics and volume averaging: {H}ow do the two techniques compare?}
  \jt{Advances in Water Resources}  \bvol{62},  \pg{178--206}.

\bibitem[Davit {\em et~al.\/}(2013{\natexlab{{\em b\/}}})Davit, Byrne, Osborne,
  Pitt-Francis, Gavaghan \& Quintard]{davit2013hydrodynamic}
{\sc \au{Davit, Y.}, \au{Byrne, H.}, \au{Osborne, J.}, \au{Pitt-Francis, J.},
  \au{Gavaghan, D.} \& \au{Quintard, M.}} \yr{2013{\natexlab{{\em b\/}}}}
  \at{Hydrodynamic dispersion within porous biofilms}.  \jt{Physical Review E}
  \bvol{87}~(1),  \pg{012718}.

\bibitem[Domenico \& Schwartz(1990)]{domenico1998physical}
{\sc \au{Domenico, P.~A.} \& \au{Schwartz, F.~W.}} \yr{1990} {\em Physical and
  Chemical Hydrogeology\/}.  \publ{Wiley, New York}.

\bibitem[Fritton \& Weinbaum(2009)]{fritton2009fluid}
{\sc \au{Fritton, S.~P.} \& \au{Weinbaum, S.}} \yr{2009}  \at{Fluid and solute
  transport in bone: flow-induced mechanotransduction}.  \jt{Annual {R}eview of
  {F}luid {M}echanics}  \bvol{41},  \pg{347--374}.

\bibitem[Hornung(1996)]{hornung1996homogenization}
{\sc \au{Hornung, U}} \yr{1996} {\em Homogenization and porous media\/}, ,
  \vol{vol.~6}.  \publ{Springer Science \& Business Media}.

\bibitem[Kuwata \& Suga(2017)]{kuwata2017direct}
{\sc \au{Kuwata, Y.} \& \au{Suga, K.}} \yr{2017}  \at{Direct numerical
  simulation of turbulence over anisotropic porous media}.  \jt{Journal of
  Fluid Mechanics}  \bvol{831},  \pg{41--71}.

\bibitem[Li {\em et~al.\/}(2018)Li, Wei, Xie, Tan, Zhang, Luo, Yuan, Song, Li,
  Shen, Ryan, Ling \& Bingqing]{li2018suppressing}
{\sc \au{Li, N.}, \au{Wei, W.}, \au{Xie, K.}, \au{Tan, J.}, \au{Zhang, L.},
  \au{Luo, X.}, \au{Yuan, K.}, \au{Song, Q.}, \au{Li, H.}, \au{Shen, C.},
  \au{Ryan, E.~M.}, \au{Ling, L.} \& \au{Bingqing, W.}} \yr{2018}
  \at{Suppressing dendritic lithium formation using porous media in lithium
  metal-based batteries}.  \jt{Nano Letters}  \bvol{18},  \pg{2067--2073}.

\bibitem[Mariani {\em et~al.\/}(2010)Mariani, Fabbri, Negrini \&
  Ribani]{mariani2010high}
{\sc \au{Mariani, G.}, \au{Fabbri, M.}, \au{Negrini, F.} \& \au{Ribani, P.~L.}}
  \yr{2010}  \at{High-{G}radient {M}agnetic {S}eparation of pollutant from
  wastewaters using permanent magnets}.  \jt{Separation and Purification
  Technology}  \bvol{72},  \pg{147--155}.

\bibitem[Mauri(1991)]{mauri1991dispersion}
{\sc \au{Mauri, R.}} \yr{1991}  \at{Dispersion, convection, and reaction in
  porous media}.  \jt{Physics of Fluids A: Fluid Dynamics}  \bvol{3}~(5),
  \pg{743--756}.

\bibitem[Mei \& Vernescu(2010)]{mei2010homogenization}
{\sc \au{Mei, C~C} \& \au{Vernescu, B}} \yr{2010} {\em Homogenization methods
  for multiscale mechanics\/}.  \publ{World scientific}.

\bibitem[Muntean \& Nikolopoulos(2020)]{muntean2020colloidal}
{\sc \au{Muntean, A} \& \au{Nikolopoulos, C}} \yr{2020}  \at{Colloidal
  transport in locally periodic evolving porous media---an upscaling exercise}.
   \jt{SIAM Journal on Applied Mathematics}  \bvol{80}~(1),  \pg{448--475}.

\bibitem[van Noorden(2009)]{van2009crystal}
{\sc \au{van Noorden, T~L}} \yr{2009}  \at{Crystal precipitation and
  dissolution in a porous medium: effective equations and numerical
  experiments}.  \jt{Multiscale Modeling \& Simulation}  \bvol{7}~(3),
  \pg{1220--1236}.

\bibitem[van Noorden \& Muntean(2011)]{van2011homogenisation}
{\sc \au{van Noorden, T~L} \& \au{Muntean, A}} \yr{2011}  \at{Homogenisation of
  a locally periodic medium with areas of low and high diffusivity}.  \jt{Eur J
  Appl Math}  \bvol{22}~(5),  \pg{493--516}.

\bibitem[{O'D}ea {\em et~al.\/}(2015){O'D}ea, Nelson, {E}l {H}aj, Waters \&
  Byrne]{o2015multiscale}
{\sc \au{{O'D}ea, R.~D.}, \au{Nelson, M.~R.}, \au{{E}l {H}aj, A.~J.},
  \au{Waters, S.~L.} \& \au{Byrne, H.~M.}} \yr{2015}  \at{A multiscale analysis
  of nutrient transport and biological tissue growth in vitro}.
  \jt{Mathematical Medicine and Biology: A Journal of the IMA}  \bvol{32}~(3),
  \pg{345--366}.

\bibitem[Olivieri {\em et~al.\/}(2020)Olivieri, Akoush, Brandt, Rosti \&
  Mazzino]{olivieri2020turbulence}
{\sc \au{Olivieri, S.}, \au{Akoush, A.}, \au{Brandt, L.}, \au{Rosti, M.~E.} \&
  \au{Mazzino, A.}} \yr{2020}  \at{Turbulence in a network of rigid fibers}.
  \jt{Physical Review Fluids}  \bvol{5},  \pg{074502}.

\bibitem[Printsypar {\em et~al.\/}(2019)Printsypar, Bruna \&
  Griffiths]{printsypar2019influence}
{\sc \au{Printsypar, G.}, \au{Bruna, M.} \& \au{Griffiths, I.~M.}} \yr{2019}
  \at{The influence of porous-medium microstructure on filtration}.
  \jt{Journal of Fluid Mechanics}  \bvol{861},  \pg{484--516}.

\bibitem[Quintard \& Whitaker(1994)]{quintard1994convection}
{\sc \au{Quintard, M.} \& \au{Whitaker, S.}} \yr{1994}  \at{Convection,
  dispersion, and interfacial transport of contaminants: {H}omogeneous porous
  media}.  \jt{Advances in Water Resources}  \bvol{17},  \pg{221--239}.

\bibitem[Ray {\em et~al.\/}(2012)Ray, van Noorden, Frank \&
  Knabner]{ray2012multiscale}
{\sc \au{Ray, N.}, \au{van Noorden, T.}, \au{Frank, F.} \& \au{Knabner, P.}}
  \yr{2012}  \at{Multiscale modeling of colloid and fluid dynamics in porous
  media including an evolving microstructure}.  \jt{Transport in Porous Media}
  \bvol{95},  \pg{669--696}.

\bibitem[Richardson \& Chapman(2011)]{richardson2011derivation}
{\sc \au{Richardson, G} \& \au{Chapman, S~J}} \yr{2011}  \at{Derivation of the
  bidomain equations for a beating heart with a general microstructure}.
  \jt{SIAM J Appl Math}  \bvol{71}~(3),  \pg{657--675}.

\bibitem[Rosti {\em et~al.\/}(2020)Rosti, Pramanik, Brandt \&
  Mitra]{rosti2020breakdown}
{\sc \au{Rosti, M.~E.}, \au{Pramanik, S.}, \au{Brandt, L.} \& \au{Mitra, D.}}
  \yr{2020}  \at{The breakdown of {D}arcy's law in a soft porous material}.
  \jt{Soft Matter}  \bvol{16},  \pg{939--944}.

\bibitem[Salles {\em et~al.\/}(1993)Salles, Thovert, Delannay, Prevors,
  Auriault \& Adler]{salles1993taylor}
{\sc \au{Salles, J.}, \au{Thovert, J.-F.}, \au{Delannay, R.}, \au{Prevors, L.},
  \au{Auriault, J.-L.} \& \au{Adler, P.~M.}} \yr{1993}  \at{Taylor dispersion
  in porous media. {D}etermination of the dispersion tensor}.  \jt{Physics of
  Fluids A: Fluid Dynamics}  \bvol{5}~(10),  \pg{2348--2376}.

\bibitem[Shipley \& Chapman(2010)]{shipley2010multiscale}
{\sc \au{Shipley, R.~J.} \& \au{Chapman, S.~J.}} \yr{2010}  \at{Multiscale
  modelling of fluid and drug transport in vascular tumours}.  \jt{Bulletin of
  Mathematical Biology}  \bvol{72},  \pg{1464--1491}.

\bibitem[Spycha{\l}a \& Starzyk(2015)]{spychala2015bacteria}
{\sc \au{Spycha{\l}a, M.} \& \au{Starzyk, J.}} \yr{2015}  \at{Bacteria in
  non-woven textile filters for domestic wastewater treatment}.
  \jt{Environmental Technology}  \bvol{36}~(8),  \pg{937--945}.

\bibitem[Tomin \& Lunati(2016)]{tomin2016investigating}
{\sc \au{Tomin, P.} \& \au{Lunati, I.}} \yr{2016}  \at{Investigating
  {D}arcy-scale assumptions by means of a multiphysics algorithm}.
  \jt{Advances in Water Resources}  \bvol{95},  \pg{80--91}.

\bibitem[Vald{\'e}s-Parada \& Alvarez-Ram{\'\i}rez(2011)]{valdes2011volume}
{\sc \au{Vald{\'e}s-Parada, F~J} \& \au{Alvarez-Ram{\'\i}rez, J}} \yr{2011}
  \at{A volume averaging approach for asymmetric diffusion in porous media}.
  \jt{J Chem Phys}  \bvol{134}~(20),  \pg{204709}.

\bibitem[Wang {\em et~al.\/}(2020)Wang, Liu, Zak \& Lennartz]{wang2020effect}
{\sc \au{Wang, M.}, \au{Liu, H.}, \au{Zak, D.} \& \au{Lennartz, B.}} \yr{2020}
  \at{Effect of anisotropy on solute transport in degraded fen peat soils}.
  \jt{Hydrological Processes}  \bvol{34}~(9),  \pg{2128--2138}.

\bibitem[Wang {\em et~al.\/}(2013)Wang, Wu, Fine, Schmulen, Hu, Godin, Zhang \&
  Liu]{wang2013ciliated}
{\sc \au{Wang, Z.}, \au{Wu, H.-J}, \au{Fine, D.}, \au{Schmulen, J.}, \au{Hu,
  Y.}, \au{Godin, B.}, \au{Zhang, J. X.~J.} \& \au{Liu, X.}} \yr{2013}
  \at{Ciliated micropillars for the microfluidic-based isolation of nanoscale
  lipid vesicles}.  \jt{Lab on a Chip}  \bvol{13},  \pg{2879--2882}.

\bibitem[Whitaker(1986)]{whitaker1986flow}
{\sc \au{Whitaker, S.}} \yr{1986}  \at{Flow in porous media {I}: {A}
  theoretical derivation of {D}arcy's law}.  \jt{Transport in Porous Media}
  \bvol{1},  \pg{3--25}.

\bibitem[Whitaker(2013)]{whitaker2013method}
{\sc \au{Whitaker, S.}} \yr{2013} {\em The Method of Volume Averaging\/}.
  \publ{Springer Science \& Business Media, B.V.}

\bibitem[Wood {\em et~al.\/}(2003)Wood, Cherblanc, Quintard \&
  Whitaker]{wood2003volume}
{\sc \au{Wood, B.~D.}, \au{Cherblanc, F.}, \au{Quintard, M.} \& \au{Whitaker,
  S.}} \yr{2003}  \at{Volume averaging for determining the effective dispersion
  tensor: closure using periodic unit cells and comparison with ensemble
  averaging}.  \jt{Water Resources Research}  \bvol{39}~(8),  \pg{1210}.

\end{thebibliography}

\end{document}